\documentclass[showpacs,amsmath,amssymb,aps,twocolumn]{revtex4-1}
\usepackage{graphics}
\usepackage{graphicx}
\usepackage{txfonts}
\usepackage{bm}   
\usepackage{amsmath,mathrsfs}
\usepackage{multirow}
\begin{document}

\title{Dynamical parton distributions from DGLAP equations with nonlinear corrections}

\author{Rong Wang$^{1,2}$}\email{email: rwang@impcas.ac.cn}
\author{Xurong Chen$^{1}$}\email{email: xchen@impcas.ac.cn}

\affiliation{
$^1$ Institute of Modern Physics, Chinese Academy of Sciences, Lanzhou 730000, China\\
$^2$ School of Nuclear Science and Technology, Lanzhou University, Lanzhou 730000, China\\
}
\date{\today}

\begin{abstract}
Determination of proton parton distribution functions is present
under the dynamical parton model assumption by applying DGLAP equations
with GLR-MQ-ZRS corrections. We provide two data sets, referred as IMParton16,
which are from two different nonperturbative inputs.
One is the naive input of three valence quarks
and the other is the input of three valence quarks with flavor-asymmetric sea components.
Basically, both data sets are compatible with the experimental
measurements at high scale ($Q^2>2$ GeV$^2$). Furthermore, our analysis shows that
the input with flavor-asymmetric sea components better reproduce
the structure functions at high $Q^2$. Generally, the obtained parton distribution functions,
especially the gluon distribution function, are the good options of inputs
for simulations of high energy scattering processes. The analysis is performed
under the fixed-flavor number scheme for $n_f=$ 3, 4, 5.
Both data sets start from very low scales around 0.07 GeV$^2$, where the
nonperturbative input is directly connected to the simple picture of quark model.
These results may shed some lights on the origin
of the parton distributions observed at high $Q^2$.
\end{abstract}
\pacs{12.38.Bx, 12.38.Qk, 13.60.Hb}

\maketitle

\section{Introduction}
\label{SecI}

Hadrons are the complex systems consisting of quarks and gluons,
which makes a long and continuous way to precisely understand the hadron structure.
Thanks to the collinear factorization theorem\cite{Factori-1,Factori-2,Factori-3}
in quantum chromodynamics (QCD), the calculation of high energy hadron collision
becomes much straightforward. The calculation is the product of the calculable hard process
and the incalculable soft part which is absorbed into
the parton distribution functions (PDFs). Although incalculable so far,
parton distribution functions are universal coefficients
which can be determined by the experiments conducted worldwide.
Moreover there are some models\cite{BagPDF-1,BagPDF-2,NJLPDF,MEMPDF}
and Lattice QCD (LQCD) calculations\cite{LattPDF-1,LattPDF-2,LattPDF-3}
which try to predict/match the PDFs of proton.
PDFs in wide kinematic ranges of $Q^2$ and $x$ is an important tool
to give some theoretical predictions of high energy hadron collisions
and simulations of expected interesting physics in modern colliders
or JLab experiments of high luminosity.

Determination of PDFs of proton attracts a lot
of interests on both theoretical and experimental sides.
To Date, the most reliable and precise PDFs data
comes from the global QCD analysis of experimental data.
There have been a lot of efforts and progresses achieved
on this issue\cite{GRV95,GRV98,MSTW08,CT10,ABM12,NNPDF,HERA-data-1,MMHT2014}.
In the global analysis, firstly, the initial parton
distributions at low scale $Q_0^2\sim$ 1 GeV$^2$,
commonly called the nonperturbative input, is parameterized
using complicated functions with many parameters.
Given the nonperturbative input, the PDFs at high $Q^2$ are predicted
by using DGLAP equations from QCD theory.
Secondly, the nonperturbative input is determined by comparing the theoretical
predictions to the experimental data measured at high scale.
This procedure is usually chosen to be the least square regression method.
Finally, PDFs in a wide kinematic range is given with the obtained
optimized nonperturbative input. Although a lot of progresses have been made,
the gluon distribution at small $x$ is still poorly estimated,
which has large uncertainties\cite{ProbeG-1,ProbeG-2}.
Even worse, the gluon distributions from different
collaborations exhibit large differences.
Gluon distribution needs to be more quantitative in terms with
a number of physics issues relating to the behavior of it \cite{DStump,GWatt,SSYu}.

PDFs at low resolution scale is always confusing since it is
in the nonperturbative QCD region. However it is related to the nucleon structure
information measured at high resolution scale. Therefore the nonperturbative input
gives some valuable information of the nucleon. Besides the powerful predictions
of the QCD theory, other fundamental rules of hadron physics should also be reflected
in the nonperturbative input. How does the PDFs relate to the simple
picture of the proton made up of three quarks?
In the dynamical parton model\cite{GRV95,GRV98,jr-1,jr-2,GluckReya},
the input contains only valence quarks, valence-like light seas and valence-like gluon,
which is consistent with the dressed constitute quark model.
All sea quarks and gluons at small $x$ are dynamically produced.
In the dynamical parton approach, the gluon and sea quark distributions are excellently
constrained by the experimental data, since there are no parametrizations for
input dynamical parton distributions. Parton radiation
is the dynamical origin of sea quarks and gluons inside the proton.
It is also worthwhile to point out that the valence-like input
and PDFs generated from it are positive. In some analysis of
MSTW\cite{MRST-1,MRST-2,MSTW07}, the negative gluon density distributions
are allowed for the nonperturbative input in order to fit
the small-$x$ behavior observed at high scale.
MMHT2014 PDFs\cite{MMHT2014} supersede the MSTW2008 PDFs, which make some changes to the
parameterisation to stop negative distribution in some region of $x$,
and include LHC, updated Tevatron data and the HERA combined H1 and ZEUS data
on the total and charm structure functions in the global analysis.

The dynamical parton model is developed and extended to even low scale around $Q_0^2\sim 0.06$
GeV$^2$ in our previous works\cite{NaiveInput,AsySeaFit}.
The naive nonperturbative input\cite{NaiveInput}
with merely three valence quarks are realized, which is the simplest input for the nucleon.
In the later research\cite{AsySeaFit}, we composed a nonperturbative input which consists of
three valence quarks and flavor-asymmetric sea components, and
extracted the flavor-asymmetric sea components from various experimental data measured
at high $Q^2$. The flavor-asymmetric sea components here refer to the sea quark distributions
generated not from the QCD evolution but from the complicated nonperturbative QCD mechanisms.
In terms of the interpretation of the nonperturbative input,
the extended dynamical parton model gives the clearest physics picture.
This work is mainly based on our previous works\cite{NaiveInput,AsySeaFit}.
The extended dynamical parton model is taken in the analysis.

DGLAP equations\cite{DGLAP-D,DGLAP-GL,DGLAP-AP} based on
parton model and perturbative QCD theory successfully and quantitatively
interpret the $Q^2$-dependence of PDFs.
It is so successful that most of the PDFs are extracted by
using the DGLAP equations up to now.
And the common way of improving the accuracy of the determined PDFs
is to apply the higher order calculations of DGLAP equations.
However there are many QCD-based evolution equations and corrections to
DGLAP equations\cite{GLR,MQ,Zhu-1,Zhu-2,Zhu-3} being worked out.
It is worthwhile to apply new evolution equations in the global analysis.
There are some pioneering works\cite{NaiveInput,AsySeaFit,EHKQS}
trying to reach this aim. In this work, DGLAP equations
with GLR-MQ-ZRS corrections are taken to do the global analysis.

The main purpose of this study is to give purely dynamical gluon
distributions ($g(x,Q_0^2)=0$), which is expected to be more reliable at small $x$.
The second purpose is to connect the quark model picture of proton
to the QCD description at high energy scale. The aim is to resolve
the origin of sea quarks and gluons at high resolution scale.
The third purpose is to understand the QCD dynamics of parton radiation
and parton recombination. We want to quantify the strength of GLR-MQ-ZRS
corrections by determining  the value of parton correlation length $R$.

The organization of this paper is as follows.
Section {\ref{SecII}} lists the experimental data we used in the analysis.
Section {\ref{SecIII}} discusses the QCD evolution equations, which is the
most important tool to evaluate the PDFs. The nonperturbative input inspired
by quark model and other nonperturbative effects are discussed in Sec. {\ref{SecIV}}.
The other details of the QCD analysis are explained in Sec. {\ref{SecV}}.
Section {\ref{SecVI}} shows the results of the global fits and the comparisons of
the obtained PDFs to experimental measurements and other widely used PDF data sets.
Section {\ref{SecVII}} introduces the IMParton package which gives the interface
of the obtained PDFs.
Finally, a simple discussion and summary is given in Sec. {\ref{SecVIII}}.

\section{Experimental data}
\label{SecII}

The deeply inelastic scattering (DIS) of charged leptons on nucleon
has been the powerful tool to study nucleon structure for a long time.
The quark structure of matter is clearly acquired by decades of measurements
starting from the late 1960s with the lepton probes interacting mainly
through the electromagnetic force. The DIS data of leptons is so important
that we include only the DIS data in this work. The structure function $F_2(x,Q^2)$
data used in this analysis are taken from SLAC\cite{SLAC-data},
BCDMS\cite{BCDMS-data}, NMC\cite{NMC-data}, E665\cite{E665-data}
and HERA (H1 and ZEUS)\cite{HERA-data-1,HERA-data-2} collaborations.

In order to make sure the data is in the deep inelastic region, and to eliminate
the contributions of nucleon resonances, two kinematic requirements shown in
Eq. (\ref{Q2W2_cuts}) are performed to select the experimental data.
\begin{equation}
Q^2>2~\text{GeV}^2, W^2>4~\text{GeV}^2.
\label{Q2W2_cuts}
\end{equation}
For the neutral-current DIS, the contribution of the Z-boson exchange
can not be neglected at high $Q^2$. Therefore we compose another kinematic
cut to reduce the influence of the Z-boson exchange contribution,
which is shown in Eq. (\ref{Q2_cut2}). The Z-boson exchange contribution is
of the order $\sim$ 1\% at $Q^2=1000~\text{GeV}^2$.
\begin{equation}
Q^2<1000~\text{GeV}^2.
\label{Q2_cut2}
\end{equation}
With these kinematic requirements, we get 469, 353, 258, 53 and 763 data points
from SLAC, BCDMS, NMC, E665 and HERA experiments respectively.

SLAC was the first to perform the fixed-target DIS experiments.
The SLAC data we used is from the reanalysis of a series of eight electron inclusive scattering
experiments conducted between 1970 and 1985. The reanalysis procedure implement
some improved treatments of radiative correction and the value of $R=\sigma_L/\sigma_T$.
The minus four-momentum transfer squared $Q^2$  of SLAC experiments
are not big ($Q^2\leq 30$ GeV$^2$), and the $x$ is mainly at large $x$ ($0.06\leq x\leq0.9$)
because of the relative low beam energy. The target mass correction (TMC) should not be ignored
for the SLAC data, because of the low $Q^2$ and large $x$.
In this work, the formula of TMC\cite{TMC-1,TMC-2} is taken as,
\begin{equation}
\begin{aligned}
F_2^{TMC}(x,Q^2)=\frac{x^2}{\xi^2r^3}F_2^{(0)}(\xi,Q^2)
+\frac{6M^2x^3}{Q^2r^4}h_2(\xi,Q^2)\\
+\frac{12M^4x^4}{Q^4r^5}g_2(\xi,Q^2), \\
h_2(\xi,Q^2)=\int_{\xi}^1 du\frac{F_2^{(0)}(u,Q^2)}{u^2}, \\
g_2(\xi,Q^2)=\int_{\xi}^1 du \int_u^1 dv \frac{F_2^{(0)}(v,Q^2)}{v^2}\\
=\int_{\xi}^1 dv (v-\xi) \frac{F_2^{(0)}(v,Q^2)}{v^2}, \\
\end{aligned}
\label{TMCFor}
\end{equation}
with $r=\sqrt{1+4x^2M^2/Q^2}$, and $\xi$ the Nachtmann variable defined as
$2x/(1+\sqrt{1+4x^2M^2/Q^2})$.
Compared to the later experiments, the uncertainties of the structure functions
and the absolute normalization of SLAC data are big.

\begin{figure}[htp]
\begin{center}
\includegraphics[width=0.43\textwidth]{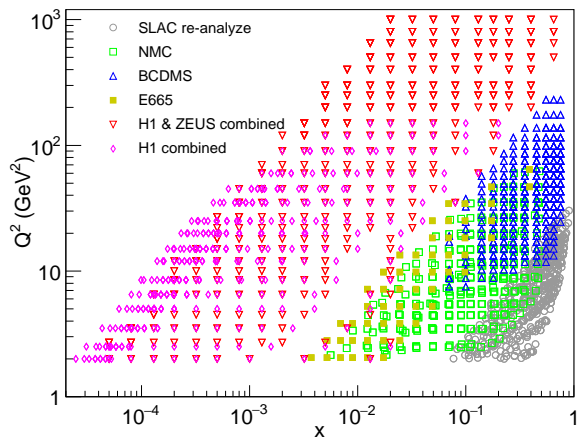}
\caption{
The kinematic coverage of the worldwide DIS data taken in the global QCD analysis.
}
\label{ExpKine}
\end{center}
\end{figure}

The precise measurements of the structure function $F_{2}$
was followed by the experiments at CERN, Fermilab, and HERA at DESY.
Both BCDMS and NMC data are collected from the muon-proton DIS
with CERN SPS muon beam but with radically different detectors.
The BCDMS data are taken at beam energy of 100, 120, 200 and 280 GeV,
and the NMC data are taken at beam energy of 90, 120, 200 and 280 GeV.
The absolute normalization for the NMC data was based on an empirical data model
motivated basically by leading order QCD calculations. Therefore we should
fit the NMC normalization factors for each incident beam energy.
The H1 and ZEUS data at HERA span a wide kinematic region of both $Q^2$
and $x$. The small $x$ information of the structure function primarily
comes from the HERA data. The HERA data we used is the combined analysis of
H1 and ZEUS experiments. The normalization uncertainty in this data is 0.5\%.
A complementary set of the inclusive HERA data was obtained by the H1 Collaboration
in the run with a reduced collision energy. These data are particularly sensitive
to the structure function $F_L$ and thereby to
the small-$x$ shape of the gluon distribution.

Finally, the kinematic coverage of the charged lepton-proton DIS data is shown
in Fig. \ref{ExpKine}. The kinematic of all the data covers 3 magnitudes in both
$x$ and $Q^2$. Since the SLAC and the NMC data distribute from relatively low
$Q^2$, the target mass corrections are applied when comparing theoretical
calculations to these data. All the normalization factors of the experimental data
are fitted in the analysis except for the combined data of H1 and ZEUS,
as the normalization uncertainties are not small for other data.

\section{Nonlinear corrections to DGLAP equations}
\label{SecIII}

DGLAP equations\cite{DGLAP-D,DGLAP-GL,DGLAP-AP} is the important and widely used
tool to describe the $Q^2$ dependence of quark and gluon densities.
The equations are derived from the perturbative QCD theory using
the quark-parton model instead of the rigorous renormalization group
equations, which offers a illuminating interpretation of the scaling violation
and the picture of parton evolution with the $Q^2$. The DGLAP equations
are written as,
\begin{equation}
\begin{aligned}
\frac{d}{dlnQ^2}
\left(
\begin{array}{c}
f_{q_i}(x,Q^2)\\
f_{g}(x,Q^2)\\
\end{array}
\right)
=\\
\Sigma_j \int_x^1\frac{dz}{z}
\left(
\begin{array}{cc}
P_{q_iq_j}(z)&P_{q_ig}(z)\\
P_{gq_j}(z)&P_{gg}(z)\\
\end{array}
\right)
\times
\left(
\begin{array}{c}
f_{q_j}(x,Q^2)\\
f_{g}(x,Q^2)\\
\end{array}
\right),
\end{aligned}
\label{DGLAP-eqs}
\end{equation}
in which $P_{q_iq_j}$, $P_{q_ig}$, $P_{gq_j}$ and $P_{gg}$ are the parton
splitting functions\cite{DGLAP-AP}. The prominent characteristic of the solution
of the equations is the rising sea quark and gluon densities toward small $x$.
The QCD radiatively generated parton distributions at small $x$ and at high $Q^2$
are tested extensively by the measurements of hard processes at modern accelerators.

The most important correction to DGLAP evolution is the parton recombination effect.
The theoretical prediction of this effect is initiated by Gribov, Levin and Ryskin (GLR)\cite{GLR},
and followed by Mueller, Qiu (MQ)\cite{MQ}, Zhu, Ruan and Shen (ZRS)\cite{Zhu-1,Zhu-2,Zhu-3}
with concrete and different methods.
The number densities of partons increase rapidly at small $x$.
At some small $x$, the number density become so large that the quanta of partons
overlap spatially. One simple criterion to estimate this saturation region
is $xf_g(x,Q^2)\geq Q^2R_p^2$, with $R_p$ the proton radius.
Therefore the parton-parton interaction effect becomes essential at small $x$,
and it expected to stop the increase of the cross sections near their unitarity limit.
In ZRS's work, the time-ordered perturbative theory (TOPT) is used
instead of the AGK cutting rules\cite{Zhu-1}. The corrections to DGLAP equations
are calculated in the leading logarithmic ($Q^2$) approximation,
and extended to the whole $x$ region, which satisfy
the momentum conservation rule\cite{Zhu-2}.

In this analysis, DGLAP equations with GLR-MQ-ZRS corrections are used to
evaluate the PDFs of proton. The GLR-MQ-ZRS corrections are very important to
slow down the parton splitting at low $Q^2<1$ GeV$^2$. Up to date, ZRS have derived
all the recombination functions for gluon-gluon, quark-gluon and quark-quark
processes\cite{Zhu-1,Zhu-2,Zhu-3}. Our previous work finds that the gluon-gluon
recombination effect is dominant\cite{AsySeaFit}, since the gluon density
is significantly larger than the quark density at small $x$.
Therefore, we use the simplified form of DGLAP equations
with GLR-MQ-ZRS corrections, which is written as,
\begin{equation}
\begin{aligned}
Q^2\frac{dxf_{q_i^{NS}}(x,Q^2)}{dQ^2}
=\frac{\alpha_s(Q^2)}{2\pi}P_{qq}\otimes f_{q_i^{NS}},
\end{aligned}
\label{ZRS-NS}
\end{equation}
for the flavor non-singlet quark distributions,
\begin{equation}
\begin{aligned}
Q^2\frac{dxf_{\bar{q}_i^{DS}}(x,Q^2)}{dQ^2}
=\frac{\alpha_s(Q^2)}{2\pi}[P_{qq}\otimes f_{\bar{q}_i^{DS}}+P_{qg}\otimes f_g]\\
-\frac{\alpha_s^2(Q^2)}{4\pi R^2Q^2}\int_x^{1/2} \frac{dy}{y}xP_{gg\to \bar{q}}(x,y)[yf_g(y,Q^2)]^2\\
+\frac{\alpha_s^2(Q^2)}{4\pi R^2Q^2}\int_{x/2}^{x}\frac{dy}{y}xP_{gg\to \bar{q}}(x,y)[yf_g(y,Q^2)]^2,
\end{aligned}
\label{ZRS-S}
\end{equation}
for the dynamical sea quark distributions, and
\begin{equation}
\begin{aligned}
Q^2\frac{dxf_{g}(x,Q^2)}{dQ^2}
=\frac{\alpha_s(Q^2)}{2\pi}[P_{gq}\otimes \Sigma+P_{gg}\otimes f_g]\\
-\frac{\alpha_s^2(Q^2)}{4\pi R^2Q^2}\int_x^{1/2} \frac{dy}{y}xP_{gg\to g}(x,y)[yf_g(y,Q^2)]^2\\
+\frac{\alpha_s^2(Q^2)}{4\pi R^2Q^2}\int_{x/2}^{x}\frac{dy}{y}xP_{gg\to g}(x,y)[yf_g(y,Q^2)]^2,
\end{aligned}
\label{ZRS-G}
\end{equation}
for the gluon distribution, in which the factor $1/(4\pi R^2)$ is from the normalization of
the two-parton densities, and $R$ is the correlation length of the two interacting partons.
In most cases, $R$ is supposed to be smaller than the hadron radius\cite{MQ}.
Note that the integral terms as $\int_x^{1/2}$ in above equations should removed
when $x$ is larger than $1/2$. $\Sigma$ in Eq. (\ref{ZRS-G}) is defined as
$\Sigma(x,Q^2)\equiv\sum_jf_{q^{NS}_j}(x,Q^2)+\sum_i[f_{q^{DS}_i}(x,Q^2)+f_{\bar{q}^{DS}_i}(x,Q^2)]$.
The splitting functions of the linear terms are given by DGLAP equations, and the recombination
functions of the nonlinear terms are written as\cite{Zhu-2},
\begin{equation}
\begin{aligned}
P_{gg\to g}(x,y)=\\
\frac{9}{64}\frac{(2y-x)(72y^4-48xy^3+140x^2y^2-116x^3y+29x^4)}{xy^5},\\
P_{gg\to q}(x,y)=\\
P_{gg\to \bar{q}}=\frac{1}{96}\frac{(2y-x)^2(18y^2-21xy+14x^2)}{y^5}.
\end{aligned}
\label{recom_funs}
\end{equation}

\section{Quark model and nonperturbative inputs}
\label{SecIV}

Quark model achieved a remarkable success in explaining the hadron spectropy
and some dynamical behaviors of high energy reactions with hadrons involved.
Quark model uncovers the internal symmetry of hadrons. Moreover,
it implies that the hadrons are composite particles containing two or three quarks.
According to the quark model assumption, the sea quarks and gluons of proton
at high $Q^2$ are radiatively produced from three valence quarks.
There are some model calculations of the initial valence quark distributions
at some low $Q_0^2$ from MIT bag model\cite{BagPDF-1,BagPDF-2},
Nambu-Jona-Lasinio model\cite{NJLPDF} and maximum entropy\cite{MEMPDF}
estimation.

Inspired by the quark model, an ideal assumption is that proton
consists of only three colored quarks at some low scale $Q_0$.
This assumption results in the naive non-perturbative input
-- three valence quarks input. At the input scale, the sea quark
and gluon distributions are all zero. This thought is widely studied
soon after the advent of QCD theory\cite{ParisiPetronzio,Vainshtein,GluckReya}.
The initial scale of the naive nonperturbative input is lower than 1 GeV$^2$,
since gluons already take comparable part of the proton energy at $Q^2=1$ GeV$^2$.
To properly evolve the naive nonperturbative input should be considered
at such low $Q^2$. Partons overlap more often at low $Q^2$
because of the big size at low resolution scale.
In our analysis, the recombination corrections are implemented.

In the dynamical PDF model, all sea quarks and gluons at small $x$
are generated by the QCD evolution processes. Global QCD analysis
based on the dynamical PDF model\cite{GRV95,GRV98,jr-1,jr-2,GluckReya}
reproduced the experimental data at high $Q^2$ with high precision
using the input of three dominated valence quarks and valence-like
components which are of small quantities.
Partons produced by the QCD evolution are called the dynamical partons.
The input scale for the valence-like input is aroud 0.3 GeV$^2$\cite{GRV95,GRV98}
and the evolution of the valence-like input is performed
with DGLAP equations. In our works, the dynamical PDF model is developed
and extended to even low $Q^2$\cite{NaiveInput,AsySeaFit}.
The naive nonperturbative input is realized in our approach.
The input of valence quarks with flavor-asymmetric sea components is also
investigated and found to be a rather better nonperturbative input.
The flavor-asymmetric sea components here refer to the intrinsic sea quarks
in the light front theory\cite{Brodsky,Chang} or
the connected sea quarks in LQCD\cite{CS1,CS2,CS3},
or the cloud sea in the $\pi$ cloud model\cite{Cloud1,Cloud2,Cloud3}.
Although there are different theories for the flavor-asymmetric sea components,
the flavor-asymmetric sea components are generated by the nonperturbative mechanisms.
These types of sea quarks are completely different from the dynamical sea quarks.
In this analysis, the evolutions of the flavor-asymmetric sea components obey
the equation for the non-singlet quark distributions.

In this work, we try to use two different inputs.
One is the naive nonperturbative input and the other is the three valence quarks
adding a few flavor-asymmetric sea components.
For convenience, three valence quarks input is called input A,
and the one with flavor-asymmetric sea components is called input B in this paper.
Accordingly, PDFs from inputs A and B are called data set A and data set B respectively.
The simplest function form to approximate valence quark distribution
is the time-honored canonical parametrization
$f(x) = A x^B (1-x)^C$, which is found to well depict
the valence distribution at large $x$.
Therefore the parameterization of the naive input is written as,
\begin{equation}
\begin{aligned}
xu^V(x,Q_0^2)=Ax^{B}(1-x)^{C},\\
xd^V(x,Q_0^2)=Dx^{E}(1-x)^{F},\\
x\bar{q}_i(x,Q_0^2)=0,\\
xg(x,Q_0^2)=0,
\end{aligned}
\label{naive-para}
\end{equation}
with zero sea quark distributions and zero gluon distribution.
One proton has two up valence quarks and one down valence quark.
Therefore we have the valence sum rules for the nonperturbtive inputs,
\begin{equation}
\begin{aligned}
\int_0^1u^V(x,Q_0^2)dx=2,\\
\int_0^1d^V(x,Q_0^2)dx=1.
\end{aligned}
\label{valence-sum}
\end{equation}
For the naive input, the valence quarks take all the momentum of proton.
Hence, we have the momentum sum rule for valence quarks in the naive input,
\begin{equation}
\int_0^1 x[u^V(x,Q_0^2)+d^V(x,Q_0^2)]dx=1.
\label{MomentumSum-1}
\end{equation}
With above constraints, there are only three free parameters
left for the parametrizations of the naive input.
The naive input (Eq. (\ref{naive-para})) is the simplest nonperturbative input
for proton, which simplifies the nucleon structure greatly.
For input B, the parametrizations of valence quarks and the valence sum rules
are the same. For simplicity, the parameterizations of the flavor-asymmetric
sea components in input B are given by,
\begin{equation}
\begin{aligned}
x\bar{d}^{AS}(x,Q_0^2)=Gx^H(1-x)^I,\\
x\bar{u}^{AS}(x,Q_0^2)=J(1-x)^Kx\bar{d}^{AS}(x,Q_0^2).
\end{aligned}
\label{asy-para}
\end{equation}
This parameterizations easily predict the $\bar{d}-\bar{u}$ difference.
The dynamical sea quark and gluon distributions are all zero for input B.
With the flavor-asymmetric sea components, the momentum sum rule
for input B is modified as follows,
\begin{equation}
\begin{aligned}
\int_0^1x\left[u^{NS}(x,Q_0^2)+d^{NS}(x,Q_0^2)\right]dx\\
=\int_0^1x[
u^{V}(x,Q_0^2)+2\bar{u}^{AS}(x,Q_0^2)\\
+d^{V}(x,Q_0^2)+2\bar{d}^{AS}(x,Q_0^2)
]dx=1.
\end{aligned}
\label{MomentumSum-2}
\end{equation}
In order to determine the quantity of the flavor-asymmetric sea components
with accuracy, the following constraint Eq. (\ref{AsyNum}) from E866
experiment\cite{E866} is taken in this analysis.
\begin{equation}
\begin{aligned}
\int_0^1 \left[ \bar{d}^{AS}(x,Q_0^2)+\bar{u}^{AS}(x,Q_0^2)\right] dx=0.118.
\end{aligned}
\label{AsyNum}
\end{equation}
Therefore, there are only 7 free parameters left for the parametrization
of input B. For better discussion on the quantity of flavor-asymmetric sea,
we define $\delta$ the momentum
fraction of the flavor-asymmetric sea components,
\begin{equation}
\begin{aligned}
\delta=\int_0^1x\left[2\bar{u}^{AS}(x,Q_0^2)+2\bar{d}^{AS}(x,Q_0^2)\right]dx.
\end{aligned}
\label{AsyFrac}
\end{equation}

One last thing about the nonperturbative input is the input scale $Q_0$.
According to the naive nonperturbative input, the momentum fraction taken by
valence quarks is one. By using QCD evolution for the second moments (momentum)
of the valence quark distributions\cite{MomentEvo}
and the measured moments of the valence quark distributions
at a higher $Q^2$\cite{GRV98}, we get the specific starting scale $Q_0=0.253$ GeV
for LO evolution (with $\Lambda_{QCD}=0.204$ GeV for $f=3$ flavors).
This energy scale is very close to the starting
scale for bag model PDFs which is $0.26$ GeV \cite{Steffens}.
In all, the initial scale $Q_0$ depends on the running coupling constant
and the experimental measurements at high $Q^2$.
We are sure that the initial scale $Q_0$ for the naive input
is close to the pole ($\Lambda_{\text{QCD}}$) of coupling constant.
In this analysis, the initial scale $Q_0$ is viewed as a free parameter
which can be determined by experimental data.

\section{QCD analysis}
\label{SecV}

The running coupling constant $\alpha_s$ and the quark masses are
the fundamental parameters of perturbative QCD. In fact these parameters
can be determined by the DIS data at high $Q^2$.
However these fundamental parameters are already determined
by a lot of experiments. Hence there is no need to let these parameters
to be free. The running coupling constant we choose is
\begin{equation}
\frac{\alpha_s(Q^2)}{4\pi}=\frac{1}{\beta_0ln(Q^2/\Lambda^2)},
\label{CouplingConst}
\end{equation}
in which $\beta_0=11-2n_f/3$ and $\Lambda^{3,4,5,6}_{LO}=204, 175, 132, 66.5$ MeV\cite{GRV98}.
For the $\alpha_s$ matchings, we take $m_c=1.4$ GeV, $m_b=4.5$ GeV, $m_t=175$ GeV.

The fixed flavor number scheme (FFNS) is used to deal with heavy quarks in this work.
In this approach, the heavy quarks ($c$, $b$ and $t$) will not be
considered as massless partons within the nucleon.
The number of active flavors $n_f$ in the DGLAP evolution
and the corresponding Wilson coefficients is fixed at $n_f=3$
(only $u$, $d$ and $s$ light quarks).
The heavy quark flavors are entirely produced perturbatively
from the initial light quarks and gluons.
The FFNS predictions agree with the DIS data with excellence\cite{GRV98,jr-1}.
In this analysis, only charm quark distribution is given,
since bottom and top distributions are trivial.
The charm quark distribution comes mainly from the gluon distribution
through the photon-gluon fusion subprocesses as $\gamma^*g\rightarrow c\overline{c}$,
$\gamma^*g\rightarrow c\overline{c}g$ and
$\gamma^*q(\overline{q})\rightarrow c\overline{c}q(\overline{q})$\cite{Witten,GluckCharm}.
The LO contribution of charm quarks to the structure function\cite{Witten,GluckCharm}
is calculated in this analysis.

The flavor-dependence of sea quarks is an interesting finding in
the nucleon structure study\cite{FlavorChang}. As discussed in Sec. {\ref{SecIII}},
the flavor-asymmetric sea components $\bar{u}^{AS}$ and $\bar{d}^{AS}$
result in the $\bar{d}-\bar{u}$ difference naturally.
As found in experiments\cite{HERMESs-1,HERMESs-2} and predicted by the LQCD\cite{CS3},
the strange quark distribution is lower than the up or down quark distribution.
In order to reflect the suppression of strange quark distribution,
the suppression ratio is applied as $\bar{s}=R(\bar{u}^{DS}+\bar{d}^{DS})/2$
with $R=0.8$\cite{AsySeaFit,CS3}. $\bar{u}^{DS}+\bar{d}^{DS}$ here
denotes the dynamical sea quarks. In this approach,
the strange quarks are all dynamical sea quarks
without any intrinsic components.

The least square method is used to determine the optimal parameterized
nonperturbative input. Using DGLAP evolution with recombination corrections,
the $\chi^2$ function is calculated by the formula,
\begin{equation}
\chi^2=\Sigma_{expt.}\Sigma_{i=1}^{N_e}\frac{(D_i-T_i)^2}{\sigma_i^2},
\label{Chi2Def}
\end{equation}
where $N_e$ is the number of data points in experiment $e$, $D_i$ is a data in a experiment,
$T_i$ is the predicted value from QCD evolution, and $\sigma_i$ is the total uncertainty
combing both statistical and systematic errors.

\section{Results}
\label{SecVI}

Two separate fits are performed for input A, which consists only three valence
quarks. One of them is the fit to all $x$ range (Fit 1) and the other
is to fit the data excluding the region of $2\times 10^{-3} <x< 0.15$ (Fit 2).
The results of the fits are listed in Table \ref{table_chi2}.
The obtained input valence quark distributions from Fit 1 and Fit 2 are expressed as
\begin{equation}
\begin{aligned}
xu^V(x,Q_0^2)=13.7x^{1.73}(1-x)^{1.55}~\text{(Fit 1)},\\
xd^V(x,Q_0^2)=6.32x^{1.13}(1-x)^{3.77}~\text{(Fit 1)},\\
xu^V(x,Q_0^2)=20.2x^{1.89}(1-x)^{1.91}~\text{(Fit 2)},\\
xd^V(x,Q_0^2)=7.85x^{1.25}(1-x)^{3.65}~\text{(Fit 2)}.\\
\end{aligned}
\label{Input_A_Fit}
\end{equation}
The initial scale $Q_0$ and the parton correlation length $R$
for parton recombination are shown in Table \ref{table_para}.
The obtained $R$ values are smaller than the proton radius,
which are consistent with the previous studies\cite{NaiveInput,MQ}.
In order to justify the importance of parton-parton recombination corrections,
we also performed a global fit using DGLAP equations without GLR-MQ-ZRS corrections
to the experimental data in the range of $x\le 2\times 10^{-3}$ or $x\ge 0.15$,
as a baseline. The obtained $\chi^2/N$ is $25349/1319=19.3$, and the input
scale is $Q_0=361$ MeV. The quality of the fit is bad if we use DGLAP equations
without parton-parton recombination corrections, because parton splitting
process only generates very steep and high parton distributions at small $x$\cite{NaiveInput}.
Parton-parton recombination corrections can not be neglected if the evolution
of PDFs starts from very low resolution scale.

\begin{table}[htp]
\centering
\caption{
The obtained $\chi^2$ of Fit 1, 2 and 3.}
\begin{tabular}{cccc}
\hline
Fit NO. & Fit range & Input & $\chi^2/N$ \\
\hline
Fit 1 & all                                    & input A & 22767/1896=12.0 \\
\multirow{2}{*}{Fit 2} & $x\le 2\times 10^{-3}$ & \multirow{2}{*}{input A} & \multirow{2}{*}{5881/1319=4.46} \\
                       &  or $x\ge 0.15$       &                          &               \\
Fit 3 & all                                    & input B & 9122/1896=4.81 \\
\hline
\end{tabular}
\label{table_chi2}
\end{table}

\begin{table}[htp]
\centering
\caption{
The obtained initial scale $Q_0$ and the correlation length $R$
for Fit 1, 2 and 3.}
\begin{tabular}{cccc}
\hline
Parameter & Fit 1 & Fit 2 & Fit 3 \\
\hline
$Q_0$ (MeV)    & 244 & 259 & 282  \\
$R$ (GeV$^{-1}$) & 4.00  & 3.98 & 3.61 \\
\hline
\end{tabular}
\label{table_para}
\end{table}

\begin{figure}[htp]
\begin{center}
\includegraphics[width=0.42\textwidth]{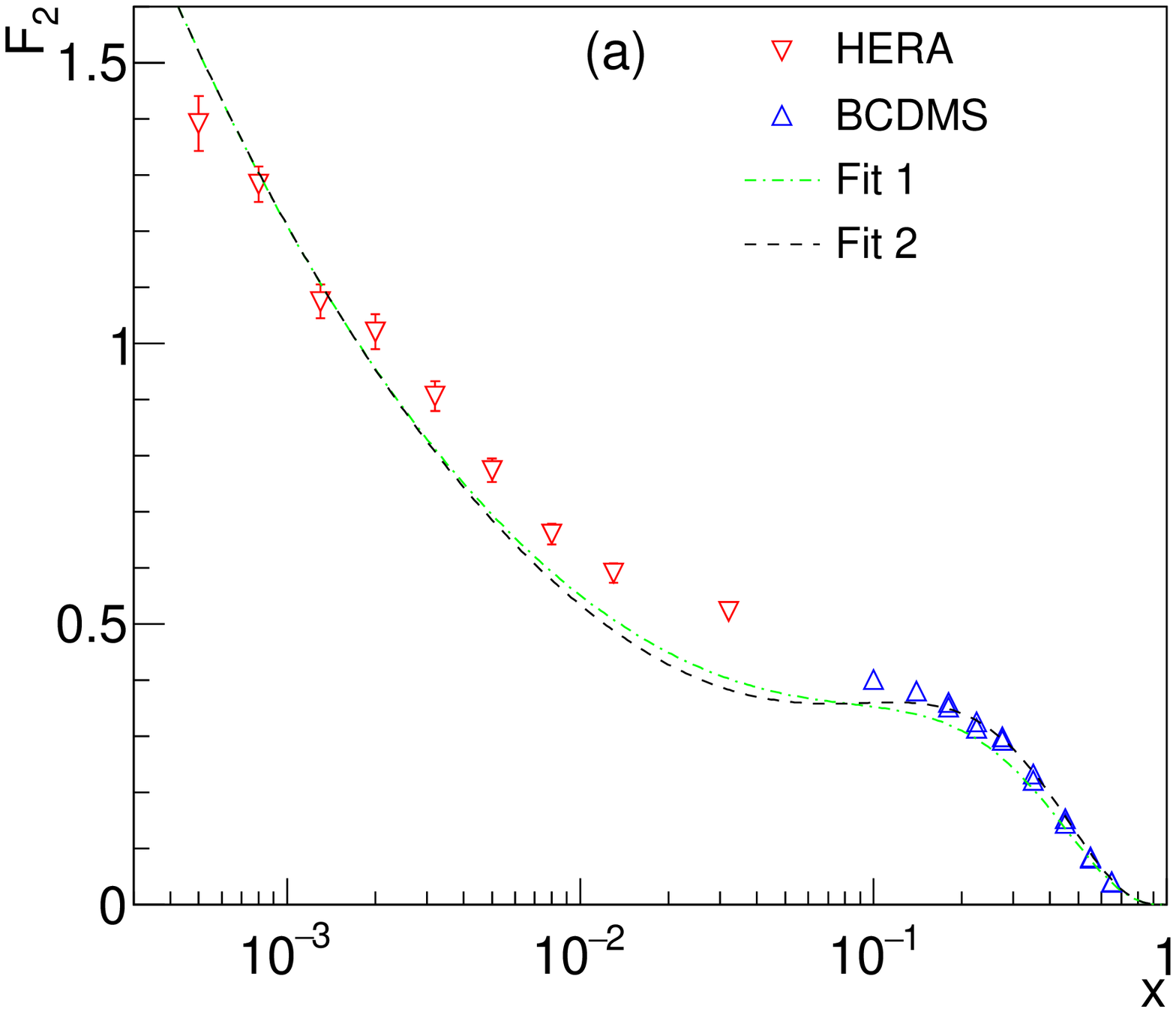}
\includegraphics[width=0.42\textwidth]{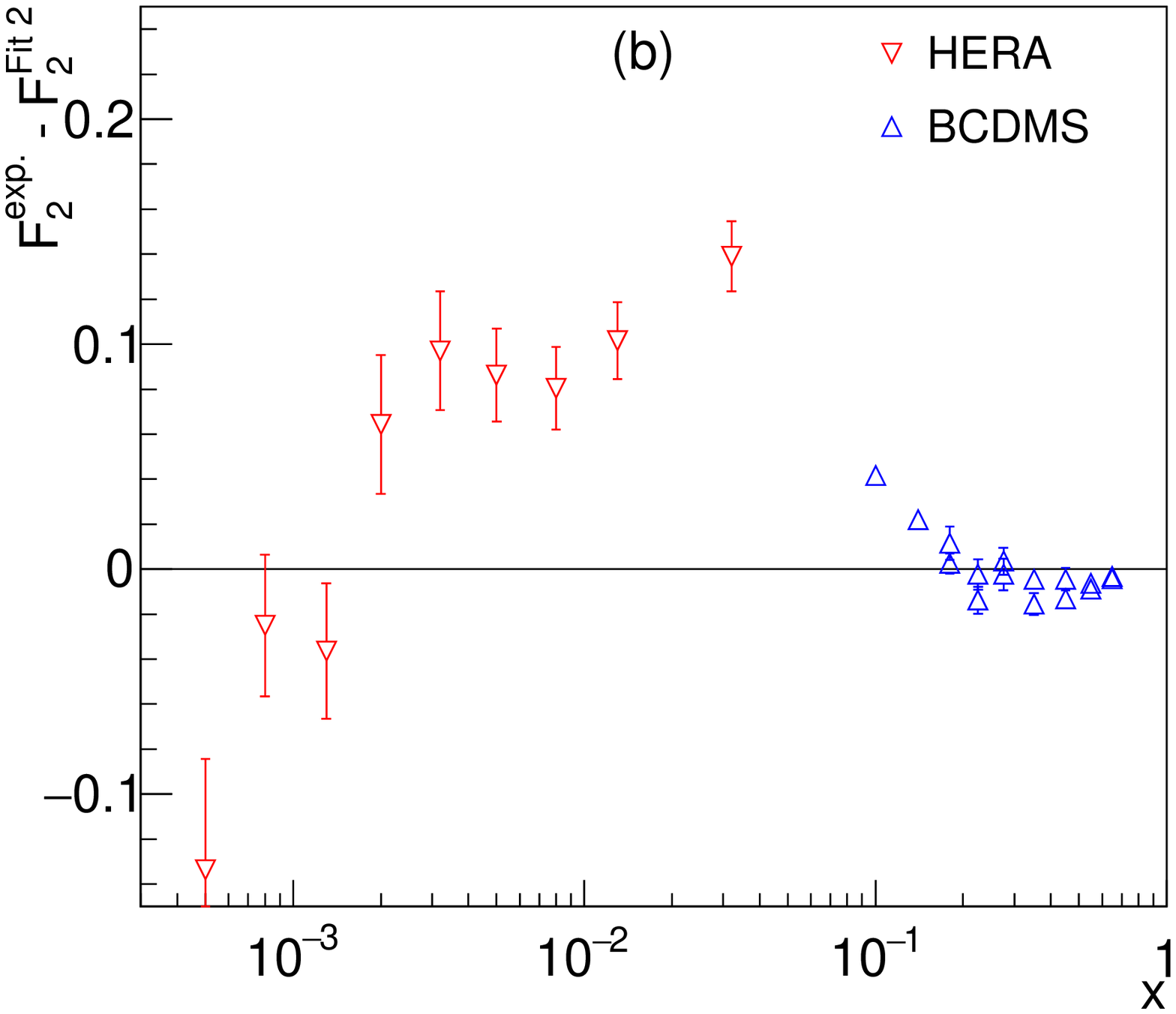}
\caption{
(a) Comparisons of the predicted $F_2$ structure functions of Fit 1 and 2
with the experimental data at $Q^2=22$ GeV$^2$\cite{HERA-data-1,BCDMS-data};
(b) The difference between the experimental data and the Fit 2 around $x=0.02$.
}
\label{Q2_22_F2}
\end{center}
\end{figure}

The obtained $\chi^2/N$ is big for input A, especially in the case
of Fit 1. Basically, the predicted $F_2$ structure function gives
the similar shape as that measured in experiments,
which are shown in Fig. \ref{Q2_22_F2}(a). However it fails in depicting
the experimental data in details around $x=0.02$.
The experimental data are obviously higher than Fit 2 in the intermediate $x$ region,
which is demonstrated clearly in Fig. \ref{Q2_22_F2}(b).
It is interesting to find that the PDFs generated from three valence quarks input
miss a peak-like component in the transition region from valence-domain to sea-domain.
Three valence quarks input needs to be modified and developed.
This discrepancy is expected to be removed by the intrinsic light quarks
or cloud sea quarks or connected quarks.

In order to get reliable valence quark distributions,
the experimental data in the region of $2\times 10^{-3} <x< 0.15$
should be excluded in the global fit, since the discrepancy around $x=0.02$
distorts the optimal three valence quarks input from the analysis.
This is the reason why we performed Fit 2 to input A.
Fit 2 is in excellence agreement with the experimental data at both large $x$
and small $x$, which are shown in Fig. \ref{Q2_22_F2}.
Quarks at small $x$ ($\lesssim 10^{-3}$) are mainly the dynamical sea quarks.
Generally, our obtained valence quark distributions and the dynamical sea quark
distributions are consistent with the experimental observables.

\begin{equation}
\begin{aligned}
xu^V(x,Q_0^2)=16.2x^{1.64}(1-x)^{2.06}~\text{(Fit 3)},\\
xd^V(x,Q_0^2)=7.45x^{1.21}(1-x)^{3.78}~\text{(Fit 3)},\\
x\bar{d}^{AS}(x,Q_0^2)=29.6x^{1.13}(1-x)^{16.8}~\text{(Fit 3)},\\
x\bar{u}^{AS}(x,Q_0^2)=1.14(1-x)^{7.11}x\bar{d}^{AS}(x,Q_0^2)~\text{(Fit 3)},\\
\end{aligned}
\label{Input_B_Fit}
\end{equation}

\begin{figure}[htp]
\begin{center}
\includegraphics[width=0.29\textwidth]{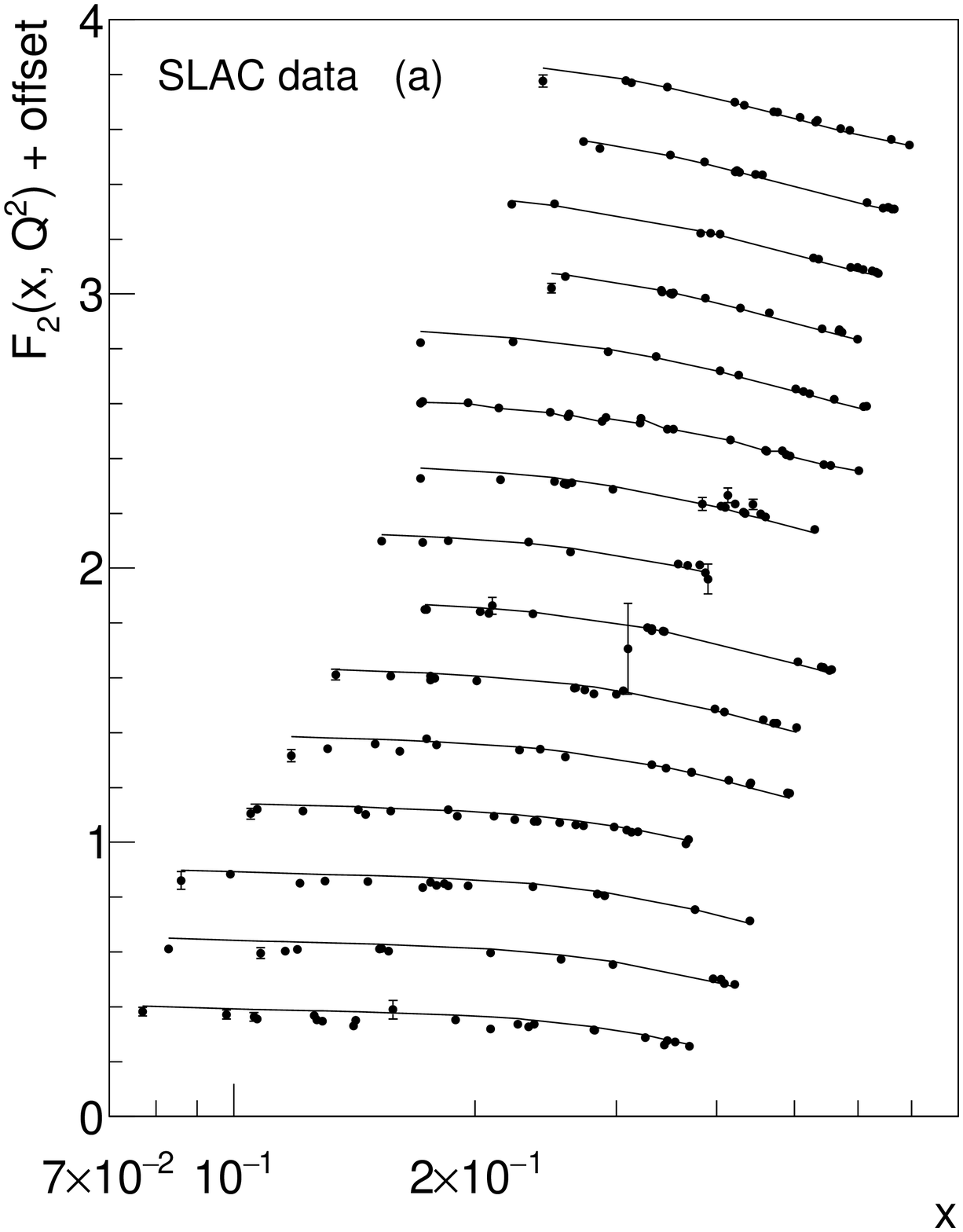}
\includegraphics[width=0.29\textwidth]{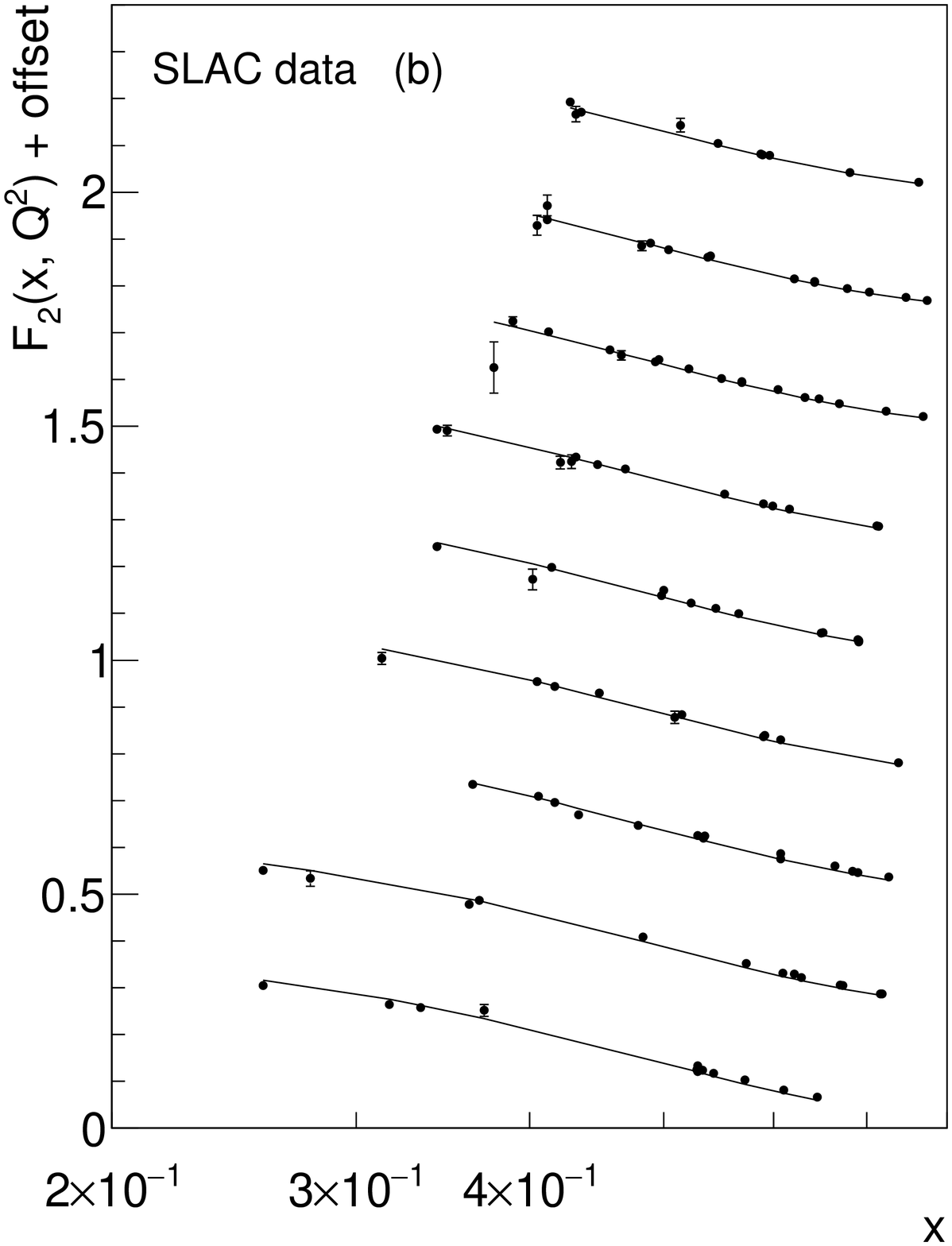}
\includegraphics[width=0.29\textwidth]{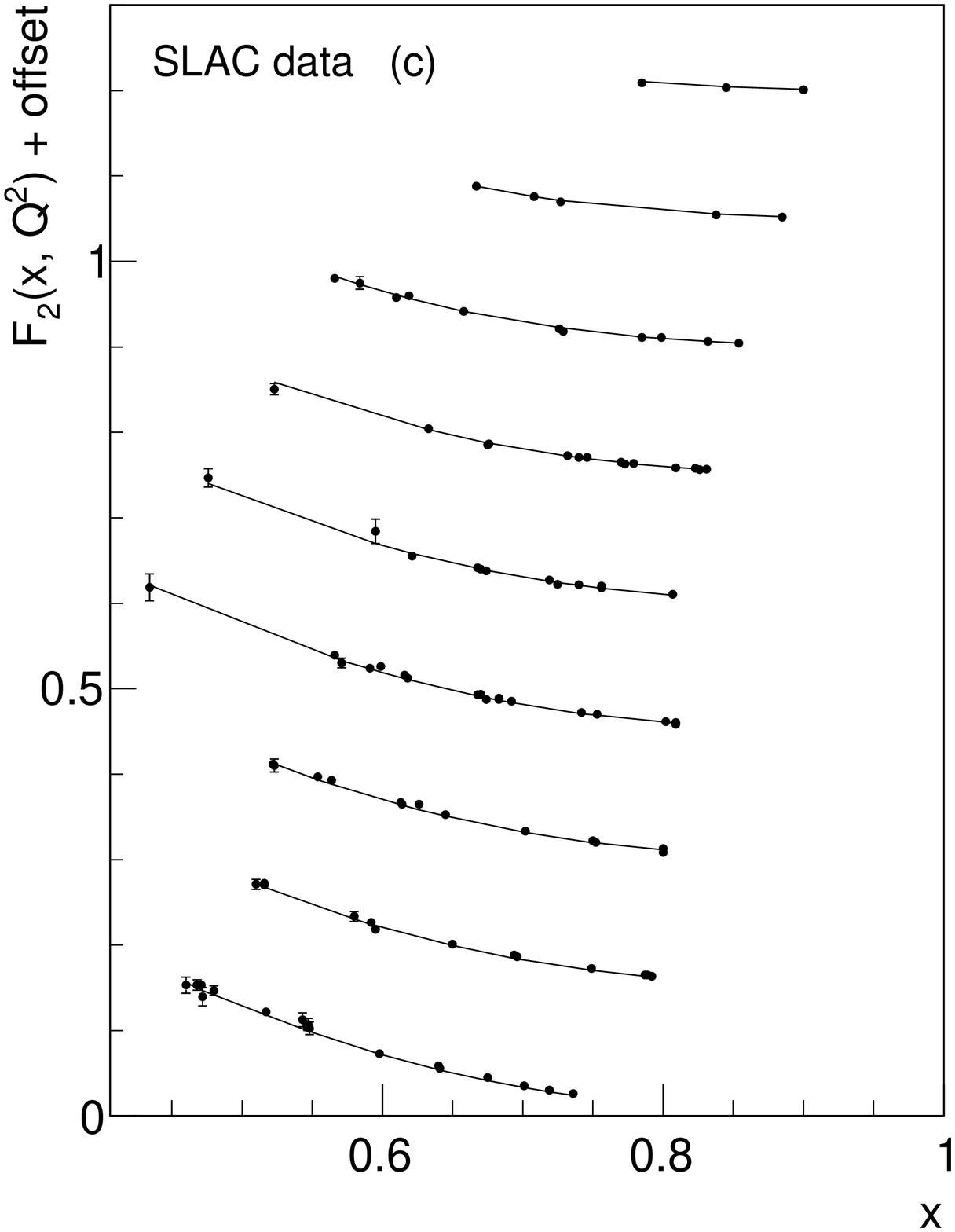}
\caption{
Comparisons of the predicted $F_2$ structure function of Fit 3 with the SLAC data\cite{SLAC-data}.
(a) From bottom to top, the mean $Q^2$ of the data are at 2.13, 2.33, 2.55, 2.93, 3.29,
3.63, 4.01, 4.28, 4.71, 5.21, 5.61, 5.98, 6.28, 6.7, and 7.21 GeV$^2$ respectively;
(b) From bottom to top, the mean $Q^2$ of the data are at 7.57, 8.01,
8.49, 9.03, 9.55, 10.2, 10.8, 11.5, and 12 GeV$^2$ respectively;
(c) From bottom to top, the mean $Q^2$ of the data are at 12.76, 13.5,
14.2, 15.3, 16.6, 18, 20.3, 23.7, and 28.5 GeV$^2$ respectively.
}
\label{SLAC_F2}
\end{center}
\end{figure}

\begin{figure}[htp]
\begin{center}
\includegraphics[width=0.31\textwidth]{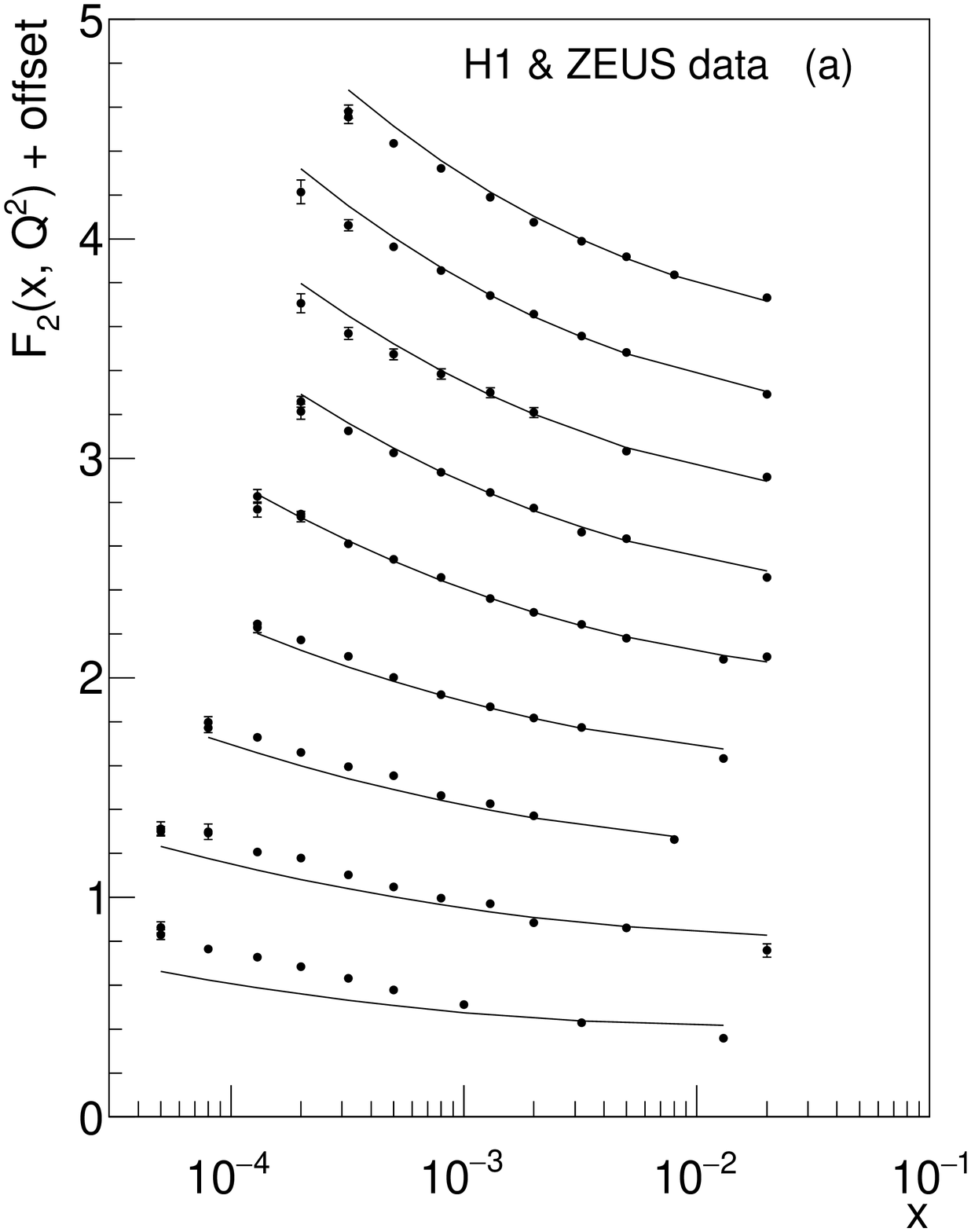}
\includegraphics[width=0.31\textwidth]{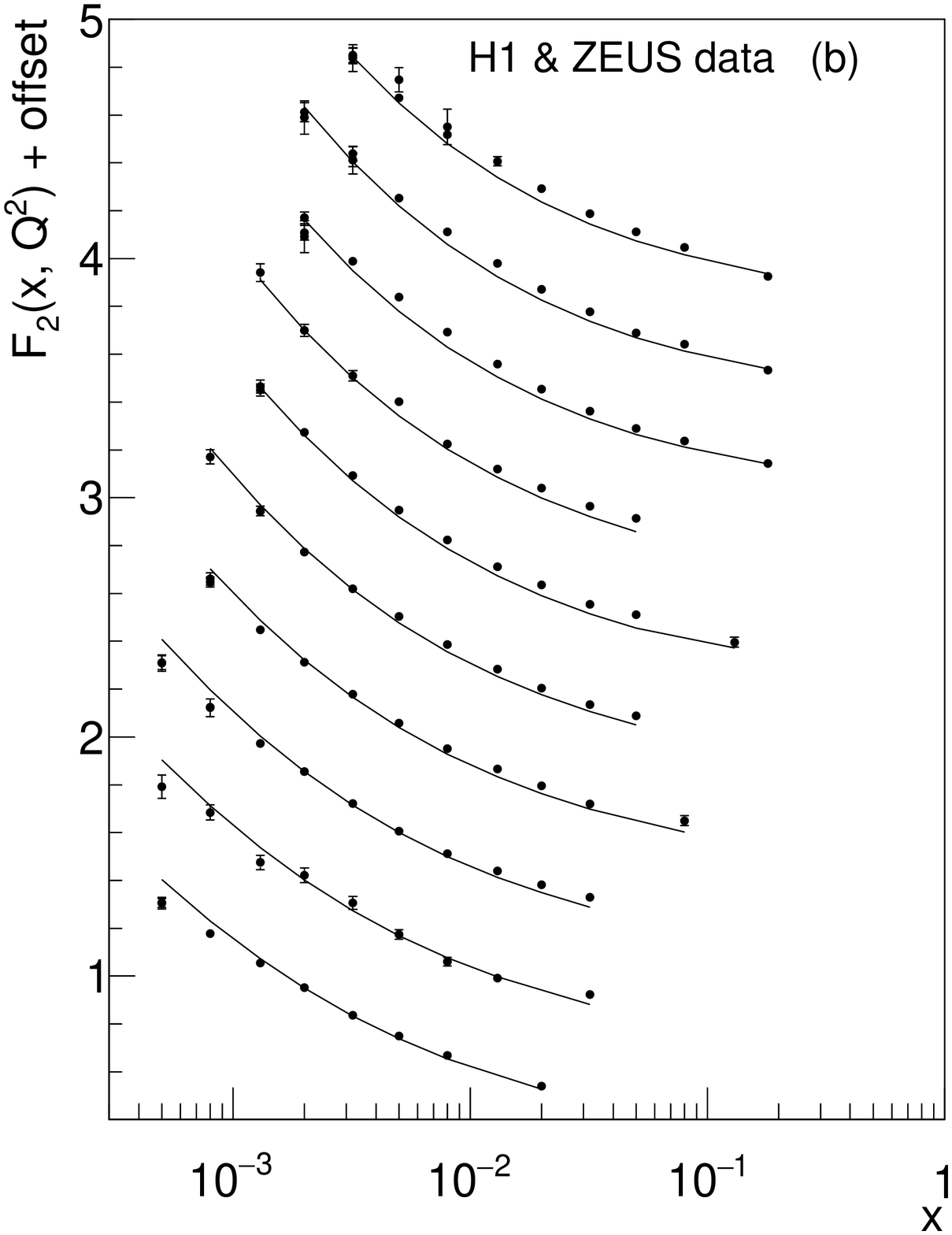}
\includegraphics[width=0.31\textwidth]{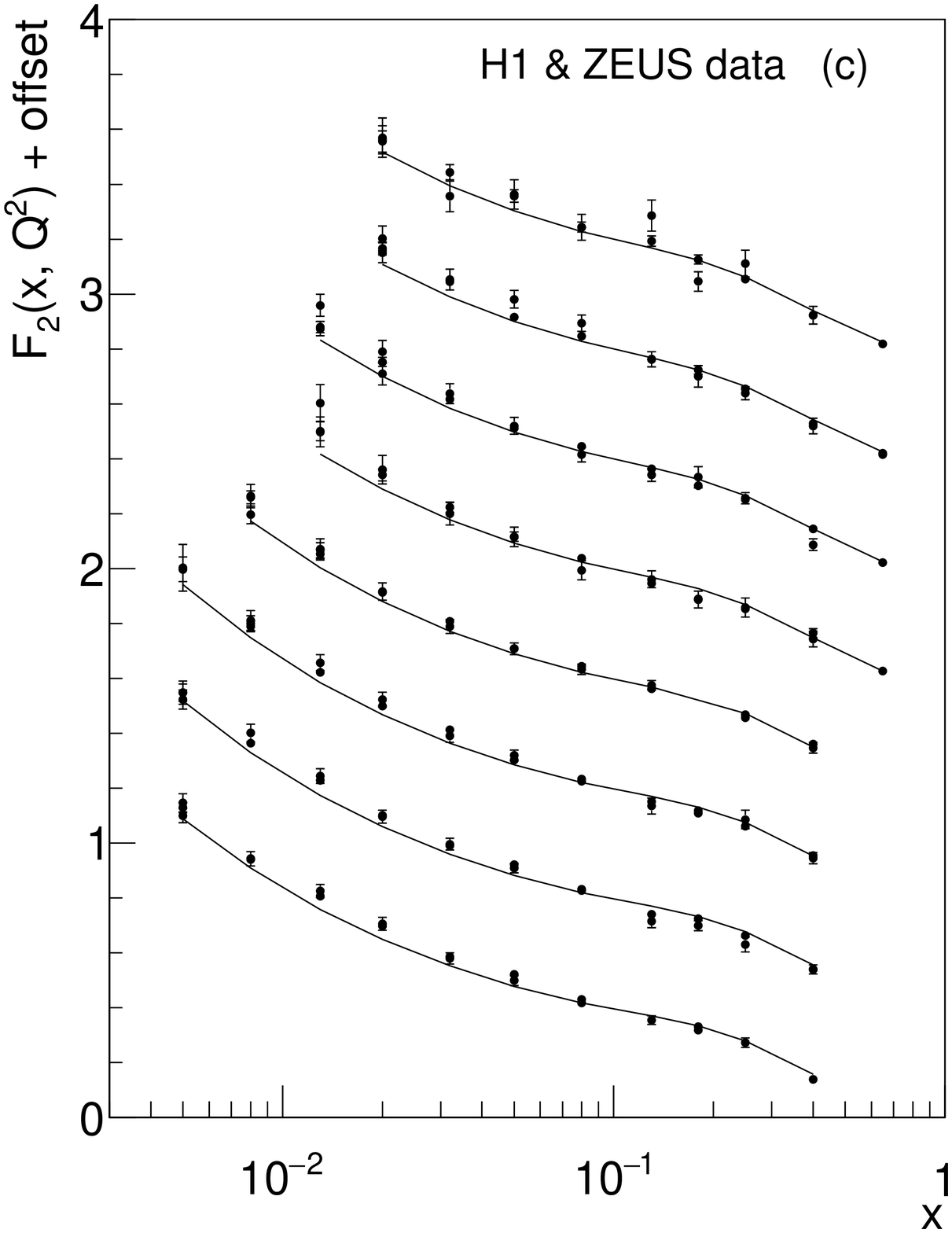}
\caption{
Comparisons of the predicted $F_2$ structure function of Fit 3 with
the combined H1 and ZEUS data\cite{HERA-data-1}.
(a) From bottom to top, the mean $Q^2$ of the data are at 2, 2.7, 3.5, 4.5, 6.5,
8.5, 10, 12, and 15 GeV$^2$ respectively;
(b) From bottom to top, the mean $Q^2$ of the data are at 18, 22,
27, 35, 45, 60, 70, 90, 120, and 150 GeV$^2$ respectively;
(c) From bottom to top, the mean $Q^2$ of the data are at 200, 250,
300, 400, 500, 650, 800, and 1000 GeV$^2$ respectively.
}
\label{HERA_F2}
\end{center}
\end{figure}

For input B, we performed a fit to the data in all $x$ range (Fit 3).
The quality of the fit improves greatly compared to input A,
which is shown in Table \ref{table_chi2}. The additional flavor-asymmetric
sea components are important to remove the discrepancy around
$x=0.02$. The obtained input is shown in Eq. (\ref{Input_B_Fit}).
So far, we have introduced the simplest parametrization for flavor-asymmetric sea components.
We argue that more complex parametrization will further improve the result.
The total momentum $\delta$ carried by flavor-asymmetric sea components
at the input scale is obtained to be 0.1. The obtained parameters $Q_0$ and $R$
are shown in Table \ref{table_para}, which is close to that of Fit 1 and Fit 2.
The determined input scales are close to the simple theoretical estimation 0.253 GeV
as discussed in Sec. {\ref{SecIV}}. The obtained normalization factor of SLAC data is
1.007. The obtained normalization factors of NMC data at beam energy of
90, 120, 200, 280 GeV are 1.07, 1.08, 1.07, and 1.04 respectively.
The normalization factors of NMC data by ABM11 global analysis\cite{ABM12}
are also large than one.
The obtained normalization factor of E665 data is 1.09.
The obtained normalization factor of H1 data is 1.02.
The obtained normalization factors of BCDMS data at beam energy of
100, 120, 200, 280 GeV are 1.02, 1.01, 1.007, and 1.01 respectively.

\begin{figure}[htp]
\begin{center}
\includegraphics[width=0.35\textwidth]{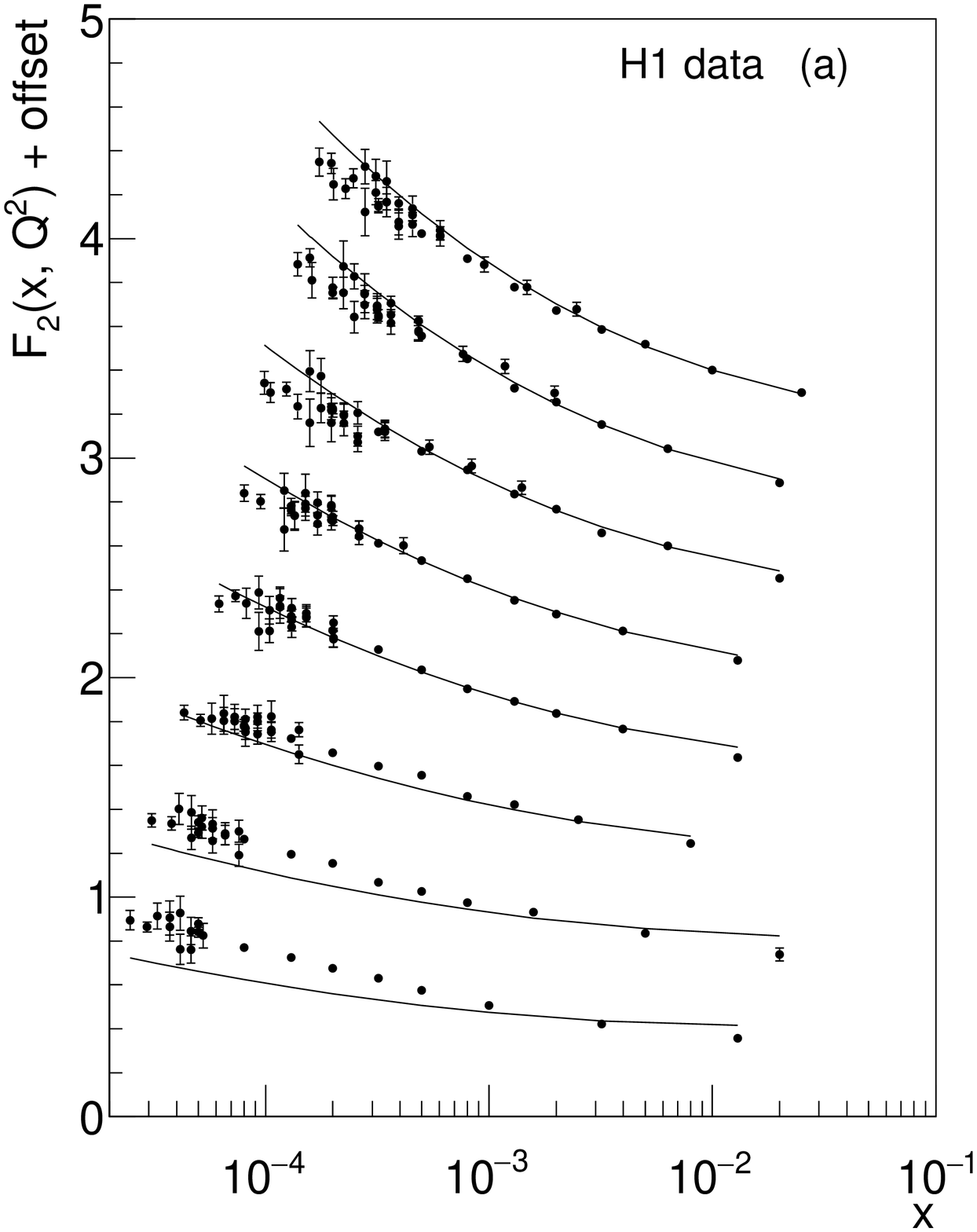}
\includegraphics[width=0.35\textwidth]{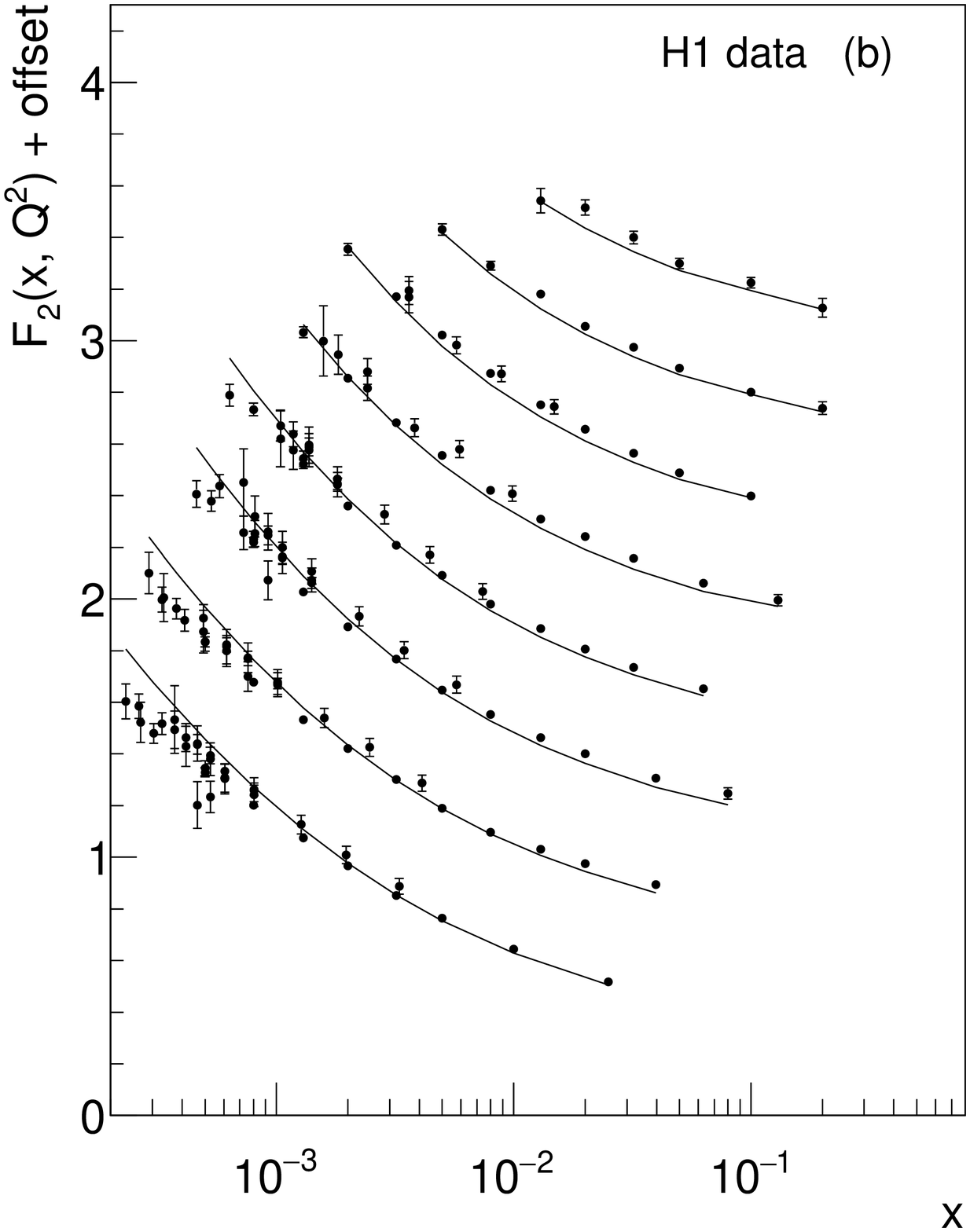}
\caption{
Comparisons of the predicted $F_2$ structure function of Fit 3 with
the H1 data\cite{HERA-data-2}.
(a) From bottom to top, the mean $Q^2$ of the data are at 2, 2.5, 3.5, 5, 6.5,
8.5, 12, and 15 GeV$^2$ respectively;
(b) From bottom to top, the mean $Q^2$ of the data are at 20, 25,
35, 45, 60, 90, 120, and 150 GeV$^2$ respectively.
}
\label{H1_F2}
\end{center}
\end{figure}

\begin{figure}[htp]
\begin{center}
\includegraphics[width=0.29\textwidth]{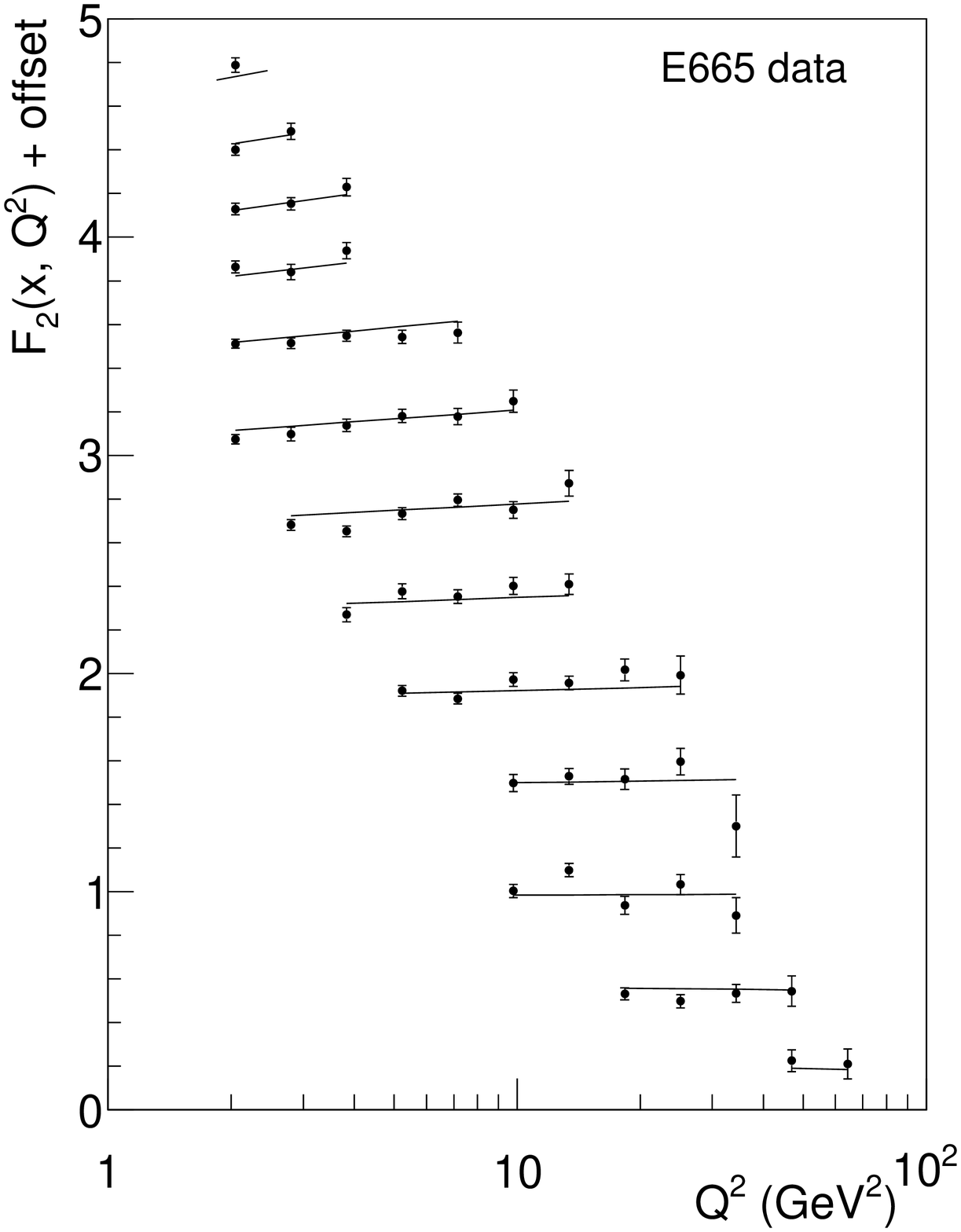}
\includegraphics[width=0.29\textwidth]{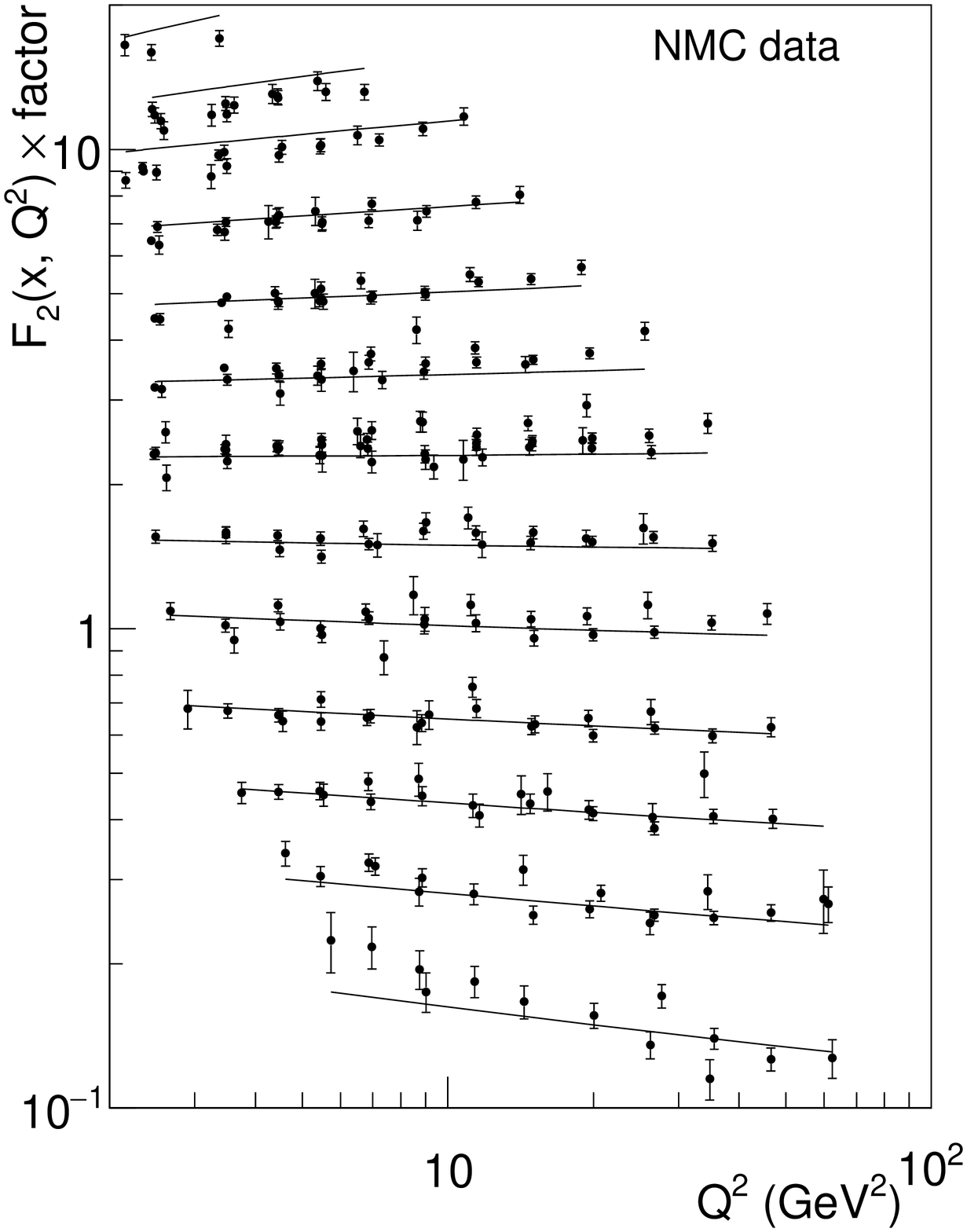}
\includegraphics[width=0.29\textwidth]{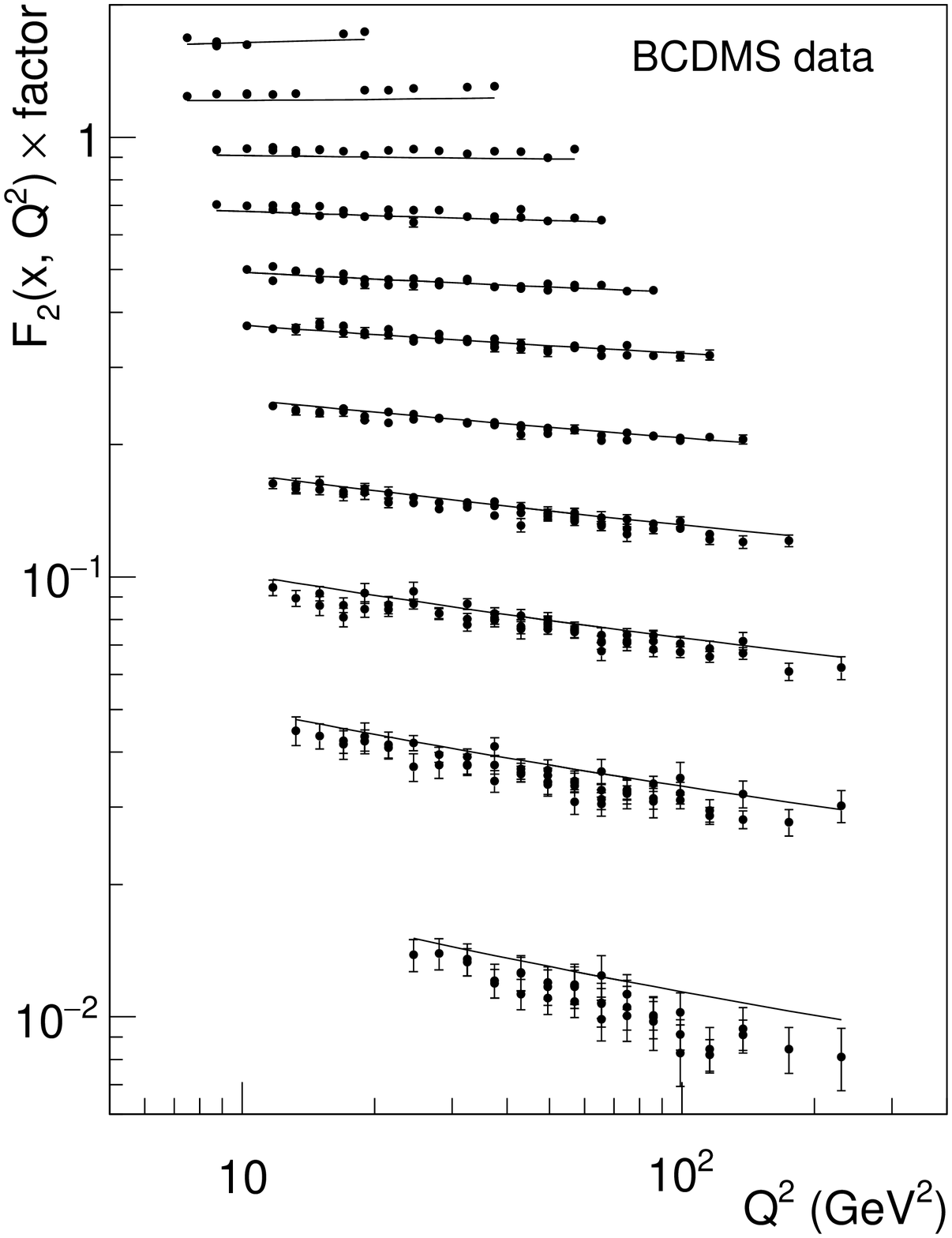}
\caption{
Comparisons of the predicted $F_2$ structure function of Fit 3 with
experimental data from E665\cite{E665-data}, NMC\cite{NMC-data}
and BCDMS\cite{BCDMS-data} experiments.
E665 data: From bottom to top, the mean $x$ of the data are 0.387, 0.173, 0.098,
0.069, 0.049, 0.0346, 0.0245, 0.0173, 0.0123, 0.00893, 0.00693, 0.0052,
and 0.0037 respectively;
NMC data: From bottom to top, the mean $x$ of the data are 0.46, 0.345, 0.275,
0.225, 0.18, 0.14, 0.1, 0.07, 0.05, 0.0353, 0.0245, 0.0155,
and 0.0075 respectively;
BCDMS data: From bottom to top, the mean $x$ of the data are 0.75, 0.65, 0.55,
0.45, 0.35, 0.275, 0.225, 0.18, 0.14, 0.1, and 0.07 respectively.
}
\label{Q2_dependence_F2}
\end{center}
\end{figure}

The predictions of $x$-dependence of structure functions at different $Q^2$
are shown in Figs. \ref{SLAC_F2}, \ref{HERA_F2} and \ref{H1_F2} with the experimental data.
Our obtained PDFs agree well with the experimental measurements
in a wide kinematical range at high resolution scale.
The evolutions of $F_2$ structure function with $Q^2$ and the comparisons with
the experimental data are shown in Fig. \ref{Q2_dependence_F2}.
Parton distribution functions generated from the the valence quarks and
the flavor-asymmetric sea components at the nonperturbative region are
consistent with the experimental measurements at high $Q^2>2$ GeV$^2$
in whole $x$ region with the application of GLR-MQ-ZRS corrections
to the standard DGLAP evolution. The experimental data favors
some intrinsic components in the nonperturbative input besides
three valence quarks.

\begin{figure}[htp]
\begin{center}
\includegraphics[width=0.43\textwidth]{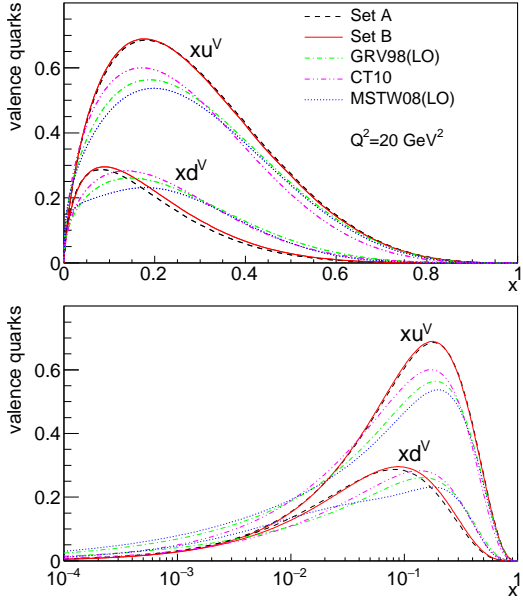}
\caption{
Comparisons of the obtained valence quark distributions with other
global QCD fits GRV98\cite{GRV98}, MSTW08\cite{MSTW08} and CT10\cite{CT10} at $Q^2=20$ GeV$^2$.
}
\label{Valence_HighQ2}
\end{center}
\end{figure}

\begin{figure}[htp]
\begin{center}
\includegraphics[width=0.43\textwidth]{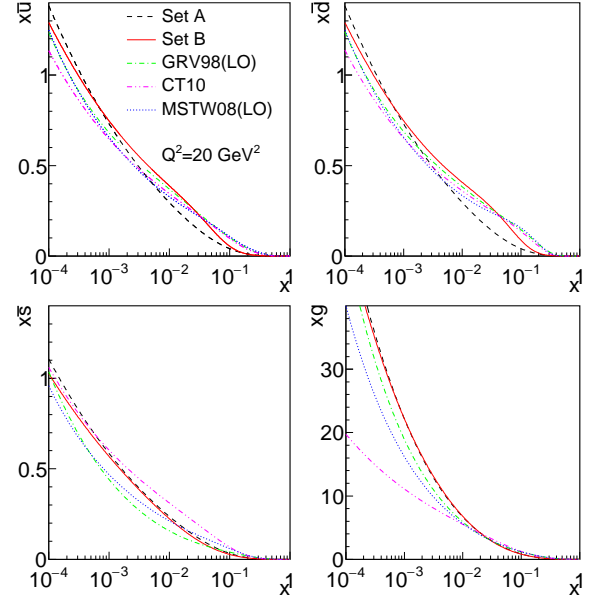}
\caption{
Comparisons of the obtained sea quark and gluon distributions with other
global QCD fits GRV98\cite{GRV98}, MSTW08\cite{MSTW08} and CT10\cite{CT10} at $Q^2=20$ GeV$^2$.
}
\label{SeaGlu_HighQ2}
\end{center}
\end{figure}

Figs. \ref{Valence_HighQ2} and \ref{SeaGlu_HighQ2} show the valence quark,
sea quark and gluon distributions at high $Q^2$
compared to other widely used parton distribution functions.
The valence quark distributions exhibit some differences between our result
and other recent global analyses. This discrepancy suggests that we
need a more complicated parametrization for valence quark distributions
beyond the the simple beta function form.
Sea quark distributions are consistent with each other.
Our gluon distribution is close to that of GRV98 and MSTW08,
but it is higher than that of CT10.
One thing we need to point out is that our gluon distributions are purely dynamically
produced in the QCD evolution. We argue that this gluon distribution is more reliable
since no arbitrary parametrization of input gluon distribution is involved.

\begin{figure}[htp]
\begin{center}
\includegraphics[width=0.43\textwidth]{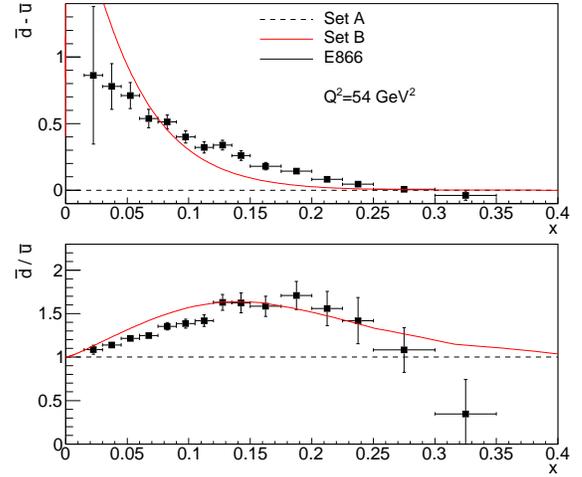}
\caption{
The predicted $\bar{u}-\bar{d}$ differences from the QCD analysis of only DIS data
are shown. The Drell-Yan data from E866 experiment\cite{E866} are also shown for comparison.
}
\label{E866_dbar_ubar}
\end{center}
\end{figure}

Our predicted difference between $\bar{d}$ and $\bar{u}$ are shown in Fig \ref{E866_dbar_ubar}.
Since the up and down dynamical sea quarks are produced from the gluon splitting,
their distributions are the same. The flavor asymmetry between up and down sea quarks
are merely from the flavor-asymmetric sea components in this approach.
The parametrization of the flavor-asymmetric sea components in this work
basically can reproduce the observed $\bar{d}$-$\bar{u}$ difference
observed in Drell-Yan process. Note that the E866 data is not included in the global
analysis.

\begin{figure}[htp]
\begin{center}
\includegraphics[width=0.43\textwidth]{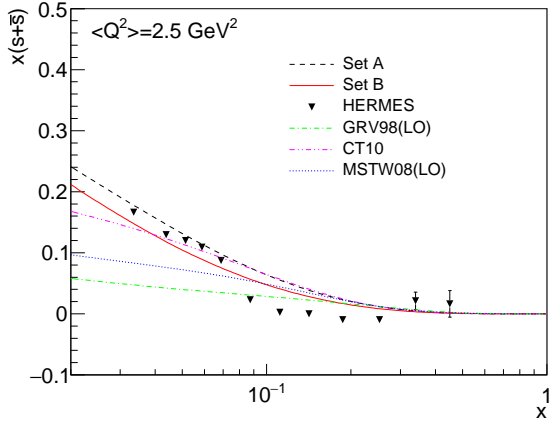}
\caption{
The obtained strange quark distribution is shown with
the experimental data from HERMES\cite{HERMESs-1,HERMESs-2} and other PDF data sets.
}
\label{Strange}
\end{center}
\end{figure}

Our predicted strange quark distribution at $Q^2=2.5$ GeV$^2$ are shown in
Fig \ref{Strange} with the recent reanalysis data by HERMES collaboration
and other widely used parton distribution functions.
The predicted strange quark distribution describes the experimental data well,
and are consistent with the other PDFs.
Our strange quark distribution is purely dynamically generated,
since there is no strange quark component in the parameterized
nonperturbative input. Compared to the up and down dynamical sea quark distributions,
the dynamical strange quark distribution is suppressed in our approach.
The suppression of the strange sea quark distribution is not hard to understand,
because the current mass of the strange quark is much heavier than that of
the up or down quark. This kind of suppression are supported by the
LQCD calculation.

\begin{figure}[htp]
\begin{center}
\includegraphics[width=0.43\textwidth]{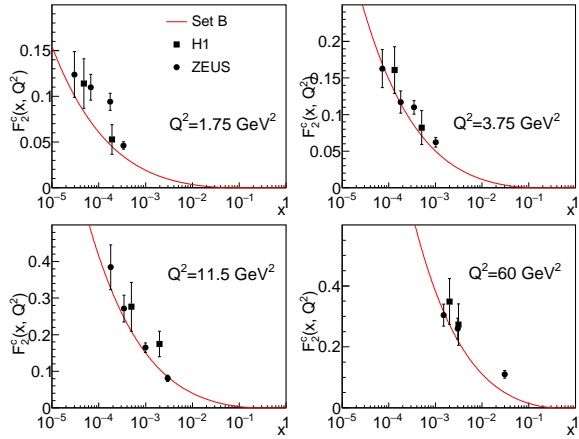}
\caption{
Comparisons of our predicted charm quark contributions
to the structure function with the experimental data
from H1\cite{F2C-H1} and ZEUS\cite{F2C-ZEUS} experiments.
}
\label{F2C}
\end{center}
\end{figure}

Fig. \ref{F2C} shows the comparison of the charm quark distributions
to the measurements by H1 and ZEUS Collaborations. The charm quark distributions are
based on LO calculation of photon-gluon fusions. This method dealing with the charm
quark distributions is also used in the global analysis by GRV95 and GRV98.
Although it is a simple calculation under the FFNS, the calculation of photon-gluon
fusion subprocesses basically reproduced the experimental measurements
of the charm quark contribution to $F_2$ structure function.

\begin{figure}[htp]
\begin{center}
\includegraphics[width=0.43\textwidth]{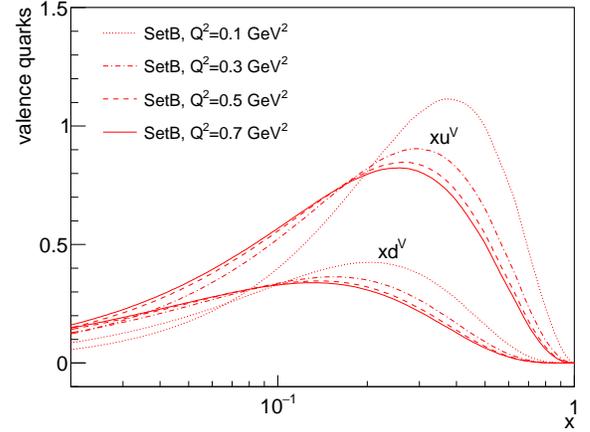}
\caption{
The predicted valence quark distributions at low resolution scales.
}
\label{Valence_at_LowQ2}
\end{center}
\end{figure}

\begin{figure}[htp]
\begin{center}
\includegraphics[width=0.43\textwidth]{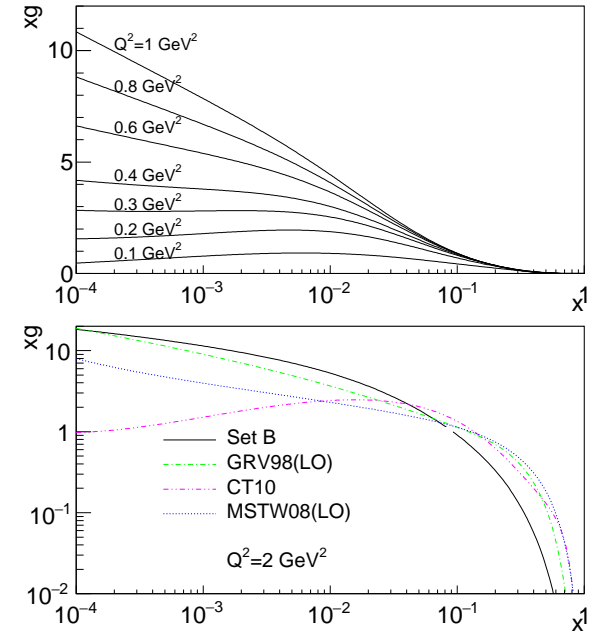}
\caption{
The predicted gluon distributions at low resolution scales.
}
\label{Gluon_at_LowQ2}
\end{center}
\end{figure}

In our approach, parton distribution functions at very low $Q^2$ are also given.
We extend the input scale form $Q_0^2=1$ GeV$^2$ down to $Q_0^2=0.1$ GeV$^2$.
Our valence quark distribution at low $Q^2$ are shown in Fig. \ref{Valence_at_LowQ2}.
The valence quark distributions are obviously high at large $x$.
Fig. \ref{Gluon_at_LowQ2} shows the gluon distributions at low $Q^2$.
The gluon distributions are Regge-like and positive at even extremely low scale.
On the issue of gluon distribution, the prominent advantage of the extended dynamical
parton model is that there is no negative gluon density
at any resolution scale, no matter how small the $Q^2$ is.

\section{IMParton package}
\label{SecVII}

We provide a C++ package named IMParton to access the obtained PDFs
in the wide kinematic range, in order to avoid the complicated QCD
evolution with GLR-MQ-ZRS corrections and make the practical applications
of the PDFs easier. The package is now available from us via email,
the WWW\cite{IMParton1.0}, or download by the git command\cite{gitIMParton}.
Two data sets of the global analysis results, called data set A (Fit 2 result)
and data set B (Fit 3 result), are provide by the package.
Data set A is from the three valence quarks nonperturbative
input, and data set B is from the nonperturbative input of three valence quarks
adding flavor-asymmetric sea quark components, as discussed in Sec. {\ref{SecIV}}.

The package consists of a C++ class named IMParton which gives the
interface to the PDFs. IMParton has a method IMParton::setDataSet(int setOption),
which let the user choose data set A or data set B via setDataSet(1)
or setDataSet(2) respectively; The most important method of IMParton is
IMParton::getPDF(int Iparton, double x, double Q2), which is the method
called by users to get the PDF values. Iparton set as -4, -3, -2, -1, 0,
1, 2, 3, 4 corresponds to getting $\bar{c}$, $\bar{s}$, $\bar{d}$, $\bar{u}$,
gluon, $u$, $d$, $s$, $c$ quark/gluon distribution functions respectively.
The given PDF values come from the quadratic interpolations of the table
grid data calculated by the DGLAP equations with GLR-MQ-ZRS corrections.
The table grids are generated in the kinematic range of $10^{-6} < x < 1$
and $0.125 < Q^2 < 2.68\times10^{8}$ GeV$^2$.  The PDF
values outside of the grid range are estimated using
some sophisticated and effective extrapolation methods.
The relative uncertainty of the interpolation is less than 1\%
in the kinematical range of $10^{-6}<x<0.9$.

\section{Discussions and summary}
\label{SecVIII}

We composed some naive nonperturbative inputs inspired by the quark model and
some other nonperturbative QCD models at very low $Q^2$.
By using DGLAP equations with GLR-MQ-ZRS corrections, PDFs generated from
these nonperturbative inputs are consistent with various experiments.
The obtained gluon distribution is purely dynamically produced, without even
the valence-like gluon distribution. The dynamical parton distributions
generated in this approach expect to have small bias
as a result of the strict theoretical constraints of the method.
A C++ package named IMParton is introduced to interface with the obtained PDFs.
Two PDF data sets are provided. One is from the three valence quarks input,
and the other is from three valence quarks with a few flavor-asymmetric sea components.
The obtained PDFs can be justified and updated with further investigations
of many other hard processes, such as the Drell-Yan process,
the inclusive jet production and the vector meson production.

By the global analysis, we find that the quark model on the proton structure
has some interesting and good results. The three valence quarks can be viewed
as the origin of the PDFs observed at high $Q^2$. Our analysis also shows that
the nonperturbatvie QCD effects beyond quark model are also needed to reproduce
the experimental data in details. By adding the flavor-asymmetric sea components
the quality of the global QCD fit improves significantly.
This is a clear evidence of the other nonperturbative parton components of proton
beyond the quark model\cite{Brodsky,Chang,CS1,CS2,CS3,Cloud1,Cloud2,Cloud3}.
It is interesting to know the fact that
the sea quarks and gluons are mainly from the parton radiations of
three valence quarks predicted by the quark model.
However there are more degrees of freedom inside proton which needs
the interpretations of the QCD theory in the future.

The nonlinear effects of parton-parton recombinations
are important at low $Q^2$ and small $x$.
Without the recombination processes, the splitting processes generate
much steep and large parton densities because of the long evolution distance
from extremely low scale. At low $Q^2$, the strength of recombination processes
are comparable to that of the parton splitting processes.
Thus the recombinations slow down enormously the fast splitting of partons
at very small $x$. The preliminary results show that the parton distribution
measurements at high $Q^2$ are directly connected to the nonperturbative
models at low scale with the applications of DGLAP equations with GLR-MQ-ZRS corrections.
DGLAP equations with nonlinear terms is a simple tool to bridge
the physics between the nonperturbative region and the perturbative region.

The last but not least conclusion we want to draw is that the partons
still exist at extremely low $Q^2$, although the definition/meaning
of the parton distribution at low scale is not clear.
The physics of partons at low $Q^2$ is affected by the parton-hadron duality,
which still needs a lot of investigations on both experimental and theoretical sides.
Based on this work, the valence quarks are the dominant partons at low $Q^2$
and go down fast at the beginning of QCD evolution (Fig. \ref{Valence_at_LowQ2}).
The dynamical sea quark and dynamical gluon distributions at low $Q^2$ and small $x$
are Regge-like, which have the flat forms over $x$ (Fig. \ref{Gluon_at_LowQ2}).
The dynamical partons grow fast at small $Q^2$ in the evolution.
The dynamical gluon distribution grows linearly with the increase of $Q^2$
instead of $ln(Q^2)$ at low $Q^2\lesssim 1$ GeV$^2$ (Fig. \ref{Gluon_at_LowQ2}).

\noindent{\bf Acknowledgments}:

We thank Wei Zhu, Fan Wang, Pengming Zhang and Jianhong Ruan
for the helpful and fruitful discussions.
One of us (R. Wang) thanks Hongkai Dai and Qiang Fu for some interesting discussions.
We are also greatly thankful to Baiyang Zhang for preparing the IMParton package.
This work was supported by the National Basic Research Program (973 Program Grant No. 2014CB845406),
and Century Program of Chinese Academy of Sciences Y101020BR0.

\end{document}